\documentclass[twocolumn,superscriptaddress,showpacs,notitlepage,prl]{revtex4-2}
\usepackage{amsmath,amssymb}
\usepackage{graphicx}
\usepackage{txfonts}
\usepackage{color}
\usepackage{hyperref}
\usepackage{adjustbox}
\usepackage{ulem}
\usepackage{bbm}
\usepackage{framed}
\usepackage{wrapfig}
\usepackage{multirow}
\usepackage[letterpaper]{geometry}
\newcommand{\be} {\begin{equation}}
\newcommand{\ee} {\end{equation}}

\global\long\def\dvec#1{\hat{#1}}

\global\long\def\Tr{{\rm Tr}}

\renewcommand{\vec}[1]{\boldsymbol{#1}}
\newcommand{\beq}{\begin{equation}}
\newcommand{\eeq}{\end{equation}}
\newcommand{\bea}{\begin{eqnarray}}
\newcommand{\eea}{\end{eqnarray}}

\global\long\def\Tr{{\rm Tr}}

\begin{document}

\title{SU(4) symmetry in twisted bilayer graphene - an itinerant perspective}

\author{Dmitry V. Chichinadze}
\affiliation{School of Physics and Astronomy,
University of Minnesota, Minneapolis, MN 55455, USA}
\author{Laura Classen}
\affiliation{Condensed Matter Physics \& Materials Science Division, Brookhaven National Laboratory, Upton, New York 11973, USA}
\author{Yuxuan Wang}
\affiliation{Department of Physics, University of Florida, Gainesville, Florida 32601, USA}
\author{Andrey V. Chubukov}
\affiliation{School of Physics and Astronomy,
University of Minnesota, Minneapolis, MN 55455, USA}

\begin{abstract}
We study symmetry-broken phases in twisted bilayer graphene at small filling above charge neutrality and at Van Hove filling. We argue that the Landau functionals for the particle-hole order parameters at these fillings both have an approximate SU(4) symmetry, but differ in the sign of quartic terms. We determine the order parameter manifold of the ground state and analyze its excitations. For small fillings, we find a strong 1st-order transition to an SU(3)$\otimes$U(1) manifold of orders that break spin-valley symmetry and induce a 3-1 splitting of fermionic excitations. For Van Hove filling, we find a weak 1st-order transition to an SO(4)$\otimes$U(1) manifold of orders that preserves the two-fold band degeneracy. We discuss the effect of particle-hole orders on superconductivity and compare with strong-coupling approaches.
\end{abstract}

\maketitle

{\it{\bf Introduction.}}~~~ Twisted bilayer graphene (TBG) is a correlated electron system near a particular ``magic'' twist angle between the layers $\theta \sim 1^\circ$, where the (quasi)periodic moire pattern with length scale of order 100 nm yields nearly flat bands separated from the rest of the energy spectrum by a gap of about $40$ meV \cite{Cao2018Insulator,Polshyn2019Linear}.
This system has attracted an enormous interest in the last few years because it displays superconductivity \cite{Cao2018SC, Yankowitz2019, Arora2020,Saito2020,Lu2019,Stepanov2020} and correlated insulating phases~\cite{Cao2018Insulator,Sharpe2019,Serlin2020,Saito2020,Saito2021Hofstadter,Wong2020,Yazdani_20,Das2021,Wu2021,Choi2021,Saito2021} near integer filling factors $|n| = 1,2,3$.

A popular theoretical approach to TBG is to treat it as a system in which Coulomb interaction well exceeds the kinetic energy~(see e.g.,  [\onlinecite{Kang2018strong,Senthil_19,Bernevig6,Khalaf2020soft,Senthil_20,Vishwanath_21,Kang2021cascades,Potasz2021exact}]  and references therein).
Within this approach, the ground states at $|n| =1,2,3$
are correlated insulators with distinct broken symmetries and band topology,
 the
  fermionic spectra
   consist of
 energy levels~\cite{Khalaf2020soft}
 or narrow sub-bands, induced by the interaction~\cite{Kang2021cascades}.

In this paper,
we discuss
 a
 complementary viewpoint, i.e. we use as the point of departure, the experimental observations~\cite{Choi_19,Yazdani_20,Wong2020,Choi2021,kitp_talk, Cao2018SC,Cao2018Insulator,Yankowitz2019,Stepanov2020,Arora2020,Wu2021,Saito2021} that in between integer fillings TBG displays metallic behavior and study how an insulator emerges from a metal as one approaches integer filling.
 In this case, at a generic non-integer
 filling $n$, fermions behave as itinerant carriers~\cite{Lin2019chiral,Chichinadze2020SC,Chichinadze2020magnet,Wang2021}, and insulating behavior near an integer $n$ emerges due to an instability in a particle-hole channel.
The corresponding order splits and reconstructs the bands and eventually drives the system into an insulating phase with narrow sub-bands.
The rationale for our approach comes from STM data~\cite{Choi_19,Yazdani_20,Wong2020,Choi2021}, which show that the density of states is non-zero everywhere in the flat region and displays Van Hove singularities, expected in the band spectrum for itinerant fermions, and from transport data, which show that the conductivity displays metallic behavior away from integer fillings \cite{kitp_talk, Cao2018SC, Cao2018Insulator,Yankowitz2019,Stepanov2020,Arora2020,Wu2021,Saito2021}.

%%%%%%%%%%%%%%%%%%%%%%%%%%%%%%%%%%%%%%%%%%%%%%%%%%%%%%%%%%%%%%
\begin{figure}[t]
\includegraphics[width=\linewidth]{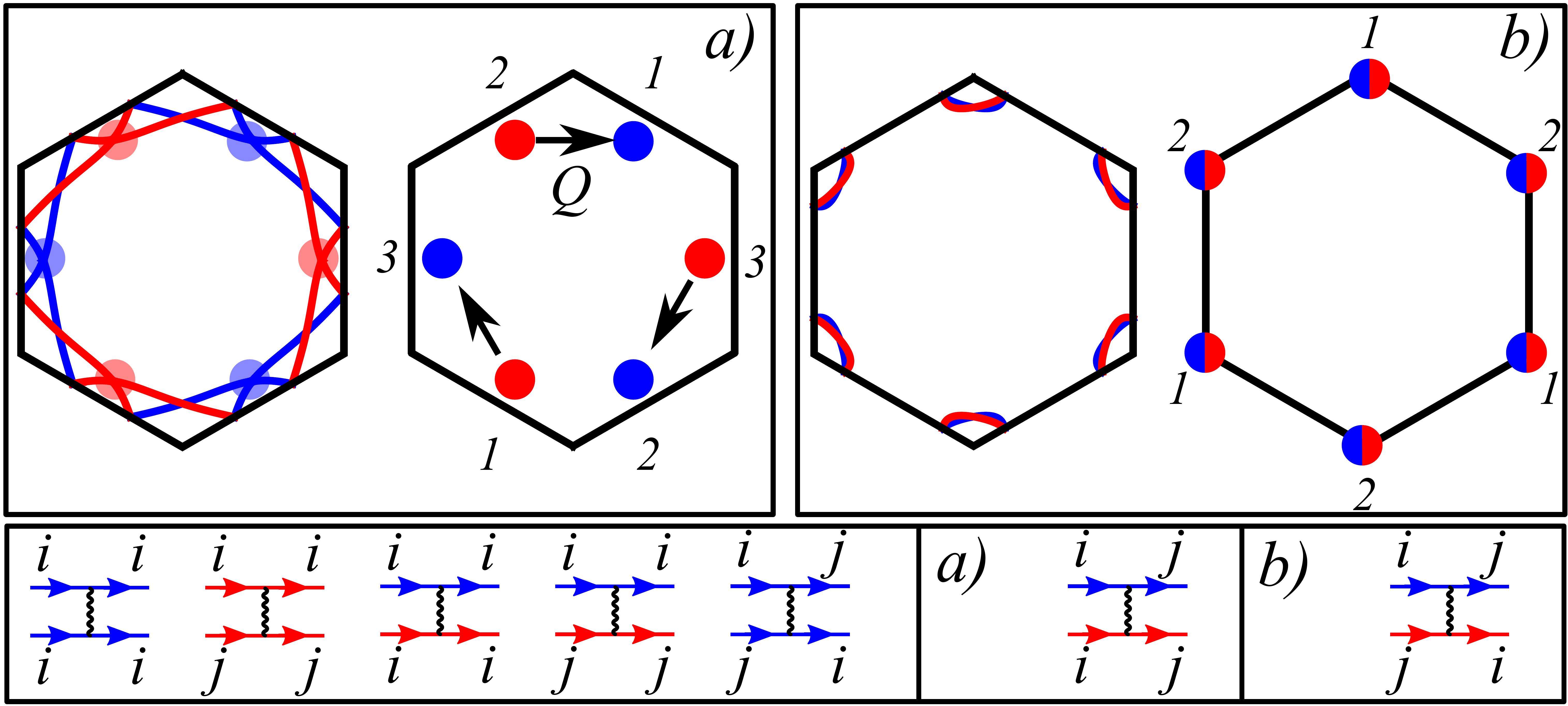}
\centering{}\caption{
Patch models and relevant interactions.
Fermions from the
two valleys are labeled by red and blue colors.
 a)
  Fermi surface and 6-patch model at Van Hove filling (color labels valley index, and $i,j$ label patches).
 b)
  same for
 the 2-patch model for
 pocketed Fermi surfaces near $K$ and $K'$.  Identical interactions
 are shown once.
 }
\label{2patch}
\end{figure}
%%%%%%%%%%%%%%%%%%%%%%%%%%%%%%%%%%%%%%%%%%%%%%%%%%%%%%%%%%%%%%

Our key results are an emergent SU(4) symmetry of itinerant fermions, which has also been argued to exist in strong-coupling approaches, and  the identification of the
manifold of degenerate ordered states, resulting from breaking of SU(4). We argue that the
manifold
 is different near different $n$. This gives rise to different degeneracies of reconstructed fermionic levels.
 We model the behavior near two exemplary $n$ by introducing patch models for typical Fermi surface geometries: pockets around the $K$, $K'$ points at small filling, and Van Hove points at intermediate filling.
We will also analyze which orders are detrimental to superconductivity and which are not.
We do not address topological properties, as the patch approximation neglects the bands that do not cross the Fermi energy.
In our case, this excludes information about Dirac points, which are at the origin of the non-trivial topological properties~\cite{Kang2018strong,PhysRevB.102.035161,PhysRevX.11.011014}.
 We conjecture that the same orders that we find based on symmetry and universal properties of the dispersion, can be extended beyond the patch approximation and give rise to proper topological behavior when added to Chern bands.

%%%%%%%%%%%%%%%%%%%%%%%%%%%%%%%%%%%%%%%%%%%%%%%%%%%%%%%%%%%%%%
\begin{figure}[h]
\includegraphics[width=0.85\linewidth]{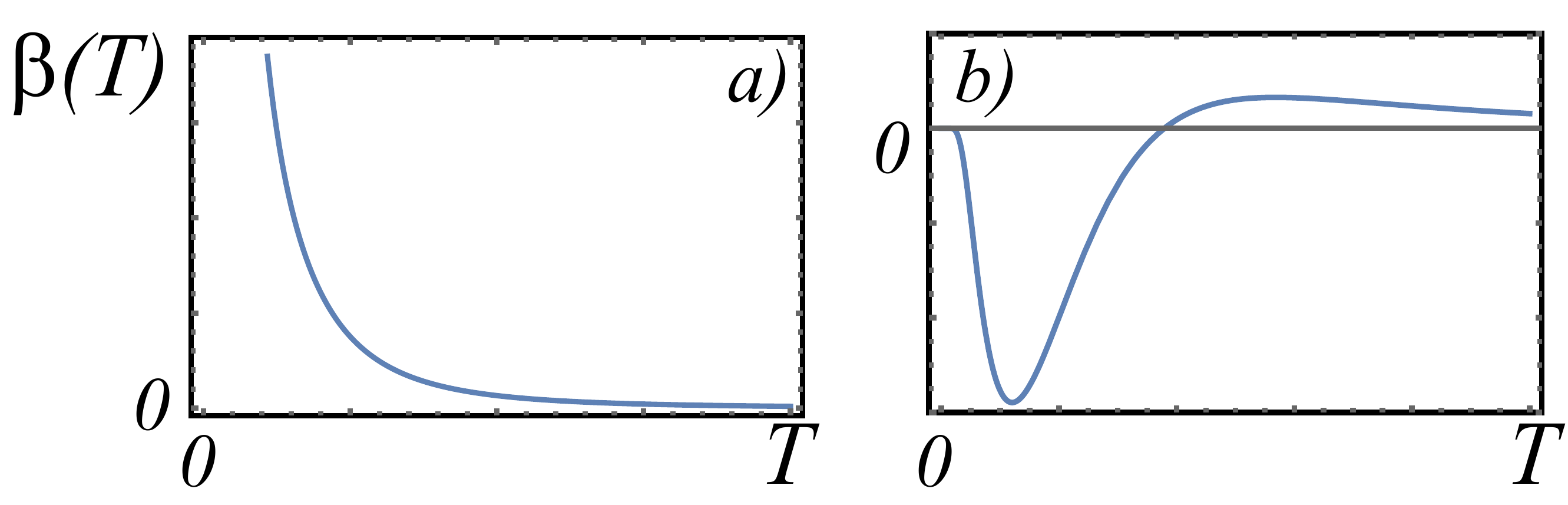}
\centering{}\caption{The prefactor $\beta$  vs $T$.
Near Van Hove filling
(a)
$\beta$ is positive for all $T$, while at smaller fillings (b)
 $\beta$ becomes negative below a certain $T$ (see \cite{SM} for the scales of $\beta$ and $T$)
 }
\label{z_plot}
\end{figure}
%%%%%%%%%%%%%%%%%%%%%%%%%%%%%%%%%%%%%%%%%%%%%%%%%%%%%%%%%%%%%%

%%%%%%%%%%%%%%%%%%%%%%%%%%%%%%%%%%%%%%%%%%%%%%%%%%%%%%%%%%%%%%
\begin{figure}[t]
\begin{minipage}[h]{0.99\linewidth}
\center{\includegraphics[width=0.9\linewidth]{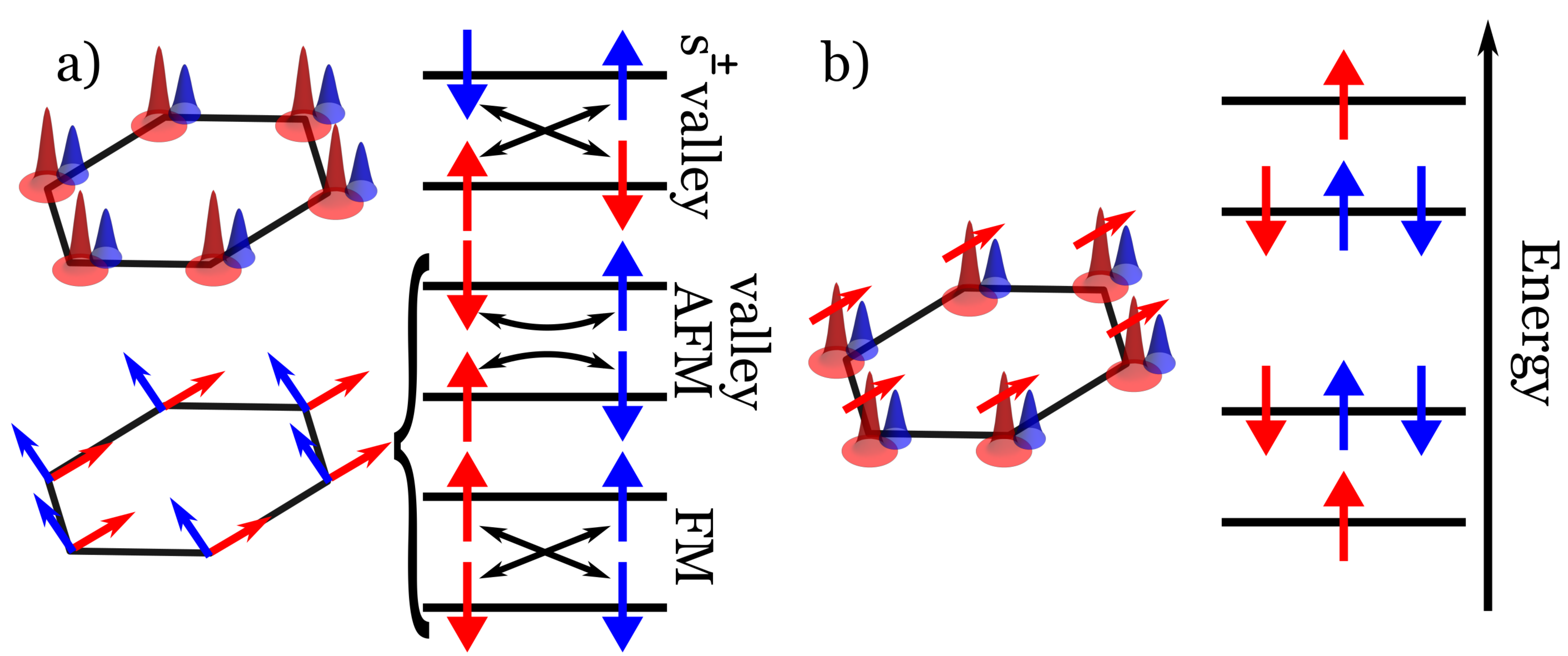} }
\end{minipage}
\centering{}\caption{
A sketch of intra-valley orders at Van Hove doping (a) and away from it (b). Left columns of panels a) and b) sketch electronic orders on the moire superlattice cell (depicted by black hexagon). Two valleys are labeled by colors (red and blue). Arrows indicate spin order for two valleys, and peaks indicate the electron density.  Right columns of panels a) and b) show the structure of energy levels for the ordered states. Double-headed arrows indicate time-reversal-partner states.}
\label{o_sketch}
\end{figure}
%%%%%%%%%%%%%%%%%%%%%%%%%%%%%%%%%%%%%%%%%%%%%%%%%%%%%%%%%%%%%%

%%%%%%%%%%%%%%%%%%%%%%%%%%%%%%%%%%%%%%%%%%%%%%%%%%%%%%%%%%%%%%
\begin{figure}[h]
\includegraphics[width=0.99\linewidth]{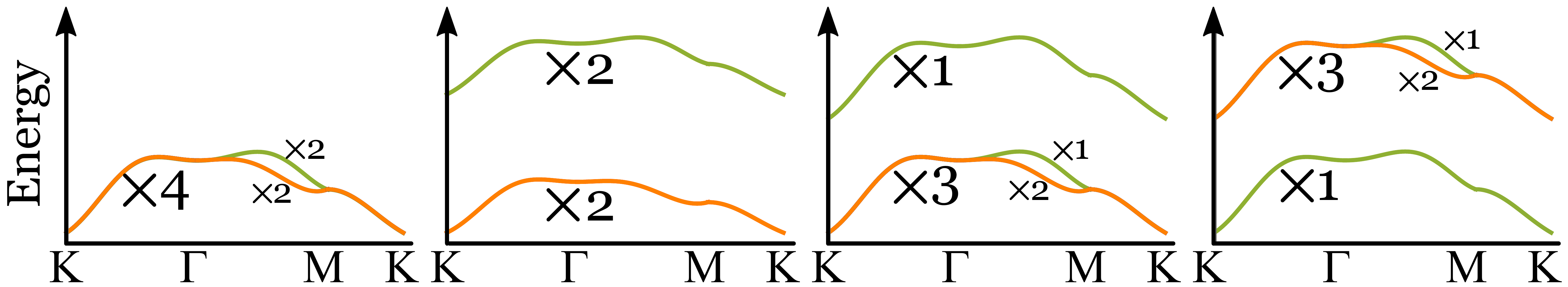}
\centering{}
\caption{Band splitting in the ground state assuming that patch orders extend to all momenta.
An almost
 4-fold
 valley and spin degenerate band splits either into two 2-fold
 degenerate bands in the 6-patch model, or into  one 3-fold
 degenerate and one non-degenerate band
in the 2-patch model.  Small $k-$regions, where this does not hold, are where the original bands are not
 valley degenerate.}
\label{band_split}
\end{figure}
%%%%%%%%%%%%%%%%%%%%%%%%%%%%%%%%%%%%%%%%%%%%%%%%%%%%%%%%%%%%%%

{\it {\bf Model.}} ~~~~The narrow spectrum of TBG contains four bands (two with positive and two with negative energy, counted from charge neutrality), each is spin-degenerate. We use the band dispersion, obtained in numerical simulations on TBG~\cite{Kang2018PRX,Yuan2018,Koshino2018PRX}, and the Kang-Vafek model~\cite{Kang2018strong,Bernevig3} for 4-fermion interactions, which includes density-density interactions and additional exchange-like interactions within a hexagon in the moire lattice.
For definiteness, we consider electron doping and focus on the two bands with positive energy. The bands are specified by the original valley index and are non-degenerate for a generic momentum. We analyze two cases: (i) Van Hove filling, when the chemical potential passes through three Van Hove points in each band
and the density of states diverges logarithmically, or even more strongly for specific band parameters\cite{Yuan2019} (the 6-patch model, Fig.~\ref{2patch}(a)) and (ii) smaller
filling, when the Fermi surface
is sizable, but still consists of pockets, centered at Dirac points $K$ and $K'$,
 (the 2-patch model, Fig.~\ref{2patch}(b)).
We
  apply the 6-patch model
   to $n \approx 2$, which experimentally is close to Van Hove filling, and the 2-patch model
    to
     fillings  around $n =1$.
    In each case we identify the set of leading particle-hole instabilities and obtain the reconstructed fermionic spectrum.

{\it{\bf SU(4) symmetry for itinerant fermions.}}~~~
A generic particle-hole order parameter $\Phi_{ij} ({\bf k}, {\bf Q})$, made out of two fermions, is specified by fermionic momenta ${\bf k}$  and ${\bf k} + {\bf Q}$ and two Pauli matrices: $\sigma_i$ acting in spin space,  and $\tau_i$ acting in ``isospin'' valley space ($i,j =0,1,2,3$, where $\sigma_0$ and $\tau_0$ are identity matrices). The effective Hamiltonian for the coupling between $\Phi_{ij} ({\bf k}, {\bf Q})$ and fermions can be cast into a $4 \times 4$ matrix form
\begin{equation}
\mathcal H_{\Phi}=
\sum_{i,j, {\bf Q}, {\bf k}} \Phi_{ij} ({\bf k},{\bf Q}) c_{\bf k}^{\dagger} \sigma_i \otimes \tau_j c_{{\bf k}+{\bf Q}},
\label{gener}
\end{equation}
where $c^{\dagger}, c$ are creation and annihilation operators of fermions.
The term with ${\bf Q} =0$ and $\sigma_0\otimes \tau_0$ can be discarded as it just renormalizes the chemical potential.
For a given filling, order parameters with
certain $\bf Q$'s
 are most likely to develop. These are, besides ${\bf Q}=0$, the various ${\bf Q}$ connecting different Van Hove points for the 6-patch model, and ${\bf Q=K-K'}$ for the 2-patch model. The ${\bf k}$ dependence can be classified by irreducible representations of the lattice point group, which are often associated with, e.g., $s$ or $d$-wave symmetry.
In the 6-patch model, the total number of components of $\Phi_{ij} ({\bf k}, {\bf Q})$ is 143 (23 for $Q=0$ and 120 for finite $Q$).
 In the 2-patch model, there are 31 fermionic bilinears with $Q=0$ and 32
 with $Q = {\bf K} - {\bf K}'$. Each order parameter gets renormalized by the interaction as $\Phi_{ij} ({\bf k}, {\bf Q}) = \Phi^{(0)}_{ij} ({\bf k}, {\bf Q})/(1- \lambda_{ij} ({\bf k}, {\bf Q}))$, where the dimensionless $\lambda_{ij}$ depends on the coupling and (temperature-dependent) susceptibility for the ordering channel. It depends on the model, which coupling(s) induce the leading instability upon lowering the temperature at $\lambda_{i,j} \rightarrow 1$.

In the 6-patch model
the two largest couplings correspond
~\cite{Chichinadze2020magnet} to a 7-component intra-valley spin and charge order ($Q=0$, $s$-wave symmetry, $i=0,...,3,j =0,3$ with $i=j=0$ excluded) and an 8-component inter-valley spin- and charge-density-wave order ($Q\neq 0$ connects neighboring Van Hove points, $s$-wave symmetry, $i=0,...,3,j =1,2$).
The two couplings are not identical, but are numerically very close for an arbitrary ratio of the density-density and the TBG-specific exchange components of the interaction. Neglecting the difference, we end up with the model of 15 order parameters specified by 15 4$\times$4 matrices  $\sigma_i \otimes \tau_j$. These 15 matrices can be viewed as orthonormal generators of an SU(4) group, and 15 corresponding order parameters form the adjoint representation of SU(4).  The free energy at the quadratic level is the sum of the squares of these 15 order parameters~\footnote{For this to hold, it is important that the corresponding generators satisfy the orthonormal condition, which in our case is $\Tr[\sigma_i \otimes \tau_j,\sigma_{i'} \otimes \tau_{j'}]=4\delta_{ii'}\delta_{jj'}$.}. We emphasize that  SU(4) is an emergent symmetry of the order parameter manifold, and the full low-energy itinerant model is not SU(4) symmetric.
A similar situation holds for the 2-patch model.
 Here we find~\cite{SM} that
    15 $Q=0$ order parameters, symmetric between patches at  $K$ and $K'$,
     have the largest and identical couplings. Neglecting other bilinears,
    we again obtain an effective model, described by 15 orthonormal generators of SU(4), with 15 order parameters forming the adjoint representation. In both models, the order parameters can be relabeled as one scalar field $\phi = \Phi_{0,3}$, two vector fields  ${\vec S}_+  =  \Phi_{i,0}$  and ${\vec S}_- =\Phi_{i,3}$,
two inter-valley scalar fields $\phi_{A,B} =\Phi_{0,j}$  and two inter-valley vector fields $\vec {S_A, S_B} = \Phi_{i,j}$
($i=1,...,3$, $j=1,2$).

{\it{\bf SU(4) Landau functional.}}~~~
To derive the Landau functional, we depart from the model of interacting fermions with
dispersion appropriate  for Van Hove  and smaller filling. We introduce
 15 order parameters in each case, use a Hubbard-Stratonovich transformation to integrate out fermions, and expand the free energy in powers of the order parameters, with coefficients evaluated using propagators of patch fermions~\cite{SM}
 \begin{align}
F= \frac{\alpha}{4}\Tr(\Phi^2) + \frac{3\gamma}{\sqrt2}\Tr(\Phi^3) + \frac{\beta}{4}\Tr(\Phi^4) + \mathcal O(\Phi^6),
\end{align}
where
$\Phi \equiv \sum_{\{i,j\}} \Phi_{ij} \sigma_i\otimes \tau_j$, and prefactors $\alpha, \beta, \gamma$ are different for the 2- and 6-patch model \footnote{
Two comments are in order. First, for a generic 6-patch model, there are three type of 4th order terms, with three coefficients  $\beta_1 = T \sum_{\omega_m} G^4_{k},  ~\beta_2 = T \sum_{\omega_m} G^3_{k} G_{k+Q}$ and $\beta_3 = T \sum_{\omega_m} G^2_{k} G^2_{k+Q}$, where $G_k = 1/(i\omega_m -\epsilon_k)$. All $\beta_i =  {\bar \beta}_i /T^2$. For the parameters of the dispersion that we use, all three ${\bar \beta}_i$ are close (see \cite{SM}), and we neglect the difference between them. In a two-patch model at $n \approx 1$, $\beta_i$ are also not identical, but differ by even smaller amount. Second, SU(4) symmetry also permits a quartic term $\propto \Tr[\Phi^2]^2$, which is absent in the microscopic derivation of $F$. It would yield independent prefactors $\beta,\beta'$ in $F^{(4)}=\beta' R^4 +4 \beta C$ in
Eq.~\eqref{Landau4}.}.
One can verify that $F$ remains invariant under $\Phi \to \Phi' = U\Phi U^\dagger$ for $U\in \rm{SU(4)}$.
Explicitly, the quadratic term has the form $F^{(2)} =   \alpha R^2$, where $R^2 = \phi^2 + \phi_{A}^2 + \phi_{B}^2 + \vec S_{+}^2 + \vec S_{-}^2 +\vec S_A^2 + \vec S_B^2 $.
The prefactor $\alpha$ is expressed via the interaction and fermionic polarization, and becomes negative below some $T_{ph}$.
The cubic term is allowed by symmetry and has the form
\begin{equation}
\begin{gathered}
F^{(3)} = 6 \sqrt{2} \gamma \,\left[ \vec S_{+} \!\cdot \left( \phi_A \vec S_A  + \phi_B \vec S_B \right) + \phi\, \vec S_+ \!\cdot \vec S_{-} + \vec S_- \!\cdot \vec S_B \times \vec S_A \right].
\end{gathered}
\label{Landau3}
\end{equation}
  The presence of
  $F^{(3)}$
  implies that the
  transition is first order. However, it is a weak first-order transition because  $\gamma \propto \int G^3$, where $G$ is the fermion propagator,
   vanishes, if we expand the dispersion to the lowest order around patch points, and we expect it to be small if we include higher-order terms.
   The
    key physics
    then
    comes from the quartic term, which is
$F^{(4)} = \beta (R^4 + 4C)$, where
\begin{widetext}
\begin{equation}
C =  \left( \vec S_{+} \cdot \vec S_{-} \right)^2 + \left( \vec S_B \times \vec S_A + \phi \vec S_{+} \right)^2 + \left( \vec S_A \times \vec S_{-} + \phi_B \vec S_{+} \right)^2 + \left( \vec S_- \times \vec S_{B} + \phi_A \vec S_{+} \right)^2 + \left( \vec S_A \cdot \vec S_{+} \right)^2 + \left( \vec S_B \cdot \vec S_{+} \right)^2 + \left(  \vec S_A \phi_A + \vec S_B \phi_B + \vec S_{-} \phi \right)^2.
\label{Landau4}
\end{equation}
\end{widetext}

The crucial difference between the 2- and 6-patch models is the sign of $\beta (T)$.  At Van Hove filling (6-patch model) $\beta (T)$ is positive and diverges as $1/T^{2}$ at $T \to 0$.
At smaller filling (2-patch model), we find $\beta (T) <0$ at relevant $T$, see Fig. \ref{z_plot}~\cite{SM}. The difference in the sign of $\beta$ has a strong impact on the type of the ordering transition and the order parameter manifold.

{\it{\bf Van Hove filling ($\beta >0$).}}~~~
In $F^{(4)}=\beta R^4+4\beta C$, the term $C$ contains the sum of full squares. For positive $\beta$, the Landau functional is then minimal if $C=0$, i.e. when the order parameters satisfy
\begin{equation}
\begin{gathered}
\left( \vec S_{+} \cdot \vec S_{-} \right) =\left( \vec S_A \cdot \vec S_{+} \right)=
 \left( \vec S_B \cdot \vec S_{+} \right)= \left(  \vec S_A \phi_A + \vec S_B \phi_B + \vec S_{-} \phi \right)=0, \\
 \left( \vec S_B \times \vec S_A + \phi \vec S_{+} \right)= \left( \vec S_A \times \vec S_{-} + \phi_B \vec S_{+} \right)=
  \left( \vec S_{-} \times \vec S_{B} + \phi_A \vec S_{+} \right) =0.
 \end{gathered}
 \label{min_cond}
\end{equation}
 For any configuration  that satisfies (\ref{min_cond}), $F= \alpha R^2 + \beta R^4$, and minimizing at $T < T_{ph}$, we obtain the non-zero value of the total $R^2 =|\alpha|/(2\beta)$.  The transition is second order without $F^{(3)}$ and weakly first order if the prefactor $\gamma$ in $F^{(3)}$ is small but finite.
 We give a general parameterization for the configurations at the minimum in \cite{SM}.
Specific examples are, e.g., configurations with only intra-valley components $\phi$ and ${\vec S}_{\pm}$ or only inter-valley components $\phi_{A,B}$ and $\vec S_{A,B}$.
For intra-valley order, there are two solutions: (i) $\phi \neq 0, {\vec S}_{\pm} =0$ and (ii) $\phi = 0$, $\vec S_{+} \cdot \vec S_{-}=0$
 with fixed ${\vec S}^2_{+} + {\vec S}^2_{-}= |\alpha|/2\beta$.
The first describes $s^{+-}$ valley order (splitting of chemical potentials for the two valleys), the second describes magnetic order with equal magnitudes of ${\vec S}_{1,2}=
(\vec S_+\pm \vec S_-)/\sqrt{2}$ in the two valleys, ${\vec S}^2_1= {\vec S}^2_2 = |a|/4\beta$, and arbitrary angle between  ${\vec S}_1$ and ${\vec S}_2$.
The two limiting cases are ferromagnetic and antiferromagnetic alignments. Ferromagnetism and  $s^{+-}$ valley order break time-reversal symmetry, but antiferromagnetism preserves it. Because Cooper pairs are formed by fermions
 from
  different valleys (see Fig. \ref{2patch}), an antiferromagnetic alignment is not detrimental to spin-singlet superconductivity, while the other two orders are.
   Configurations with only inter-valley components describe density-waves and loop-currents
    ~\cite{Chichinadze2020magnet}.
 For a generic order parameter that satisfies $F^{(4)} = \beta R^4$, nine variables remain undetermined by Eq.~\eqref{min_cond}. Because the total $R^2$ is fixed, there are 8 Goldstone modes. This can be also seen by noticing that the SU(4) symmetry is broken down to SO(4)$\otimes$U(1).  The broken symmetry
   is described by the coset SU(4)/[SO(4)$\otimes$U(1)] with 15-6-1=8 generators, which are the 8 Goldstone modes~\cite{SM}.

{\it{\bf Smaller filling ($\beta <0$).}}~~~ For negative $\beta$, the order parameter manifold is different as now one has to find configurations that {\it maximize} $C$ in (\ref{Landau4}). To get a first insight, consider a configuration with only intra-valley orders $\phi$ and ${\vec S}_{\pm}$.
A straightforward analysis shows that in this case $F^{(4)} = -\frac{7}{3} |\beta| R^4 + |\beta| {\tilde C}$, where
\begin{equation}
{\tilde C} = \left({\vec S}^2_+ - {\vec S}^2_-\right)^2 + \frac{1}{3} \left(2 \phi^2 - {\vec S}^2_+-{\vec S}^2_-\right)^2 + 4 \left[{\vec S}_+ \times {\vec S}_-\right]^2
\label{cond2}
\end{equation}
The minimum of $F^{(4)}$  is reached when  ${\tilde C}=0$, which holds when density and spin valley orders are both non-zero: $\phi =|{\vec S}_+|=|{\vec S}_-|$ and ${\vec S}_+ =\pm  {\vec S}_-$. The last condition implies that the spin order now develops only in one valley, see Fig. \ref{o_sketch}. The transition is first order, and to get the equilibrium value of $R^2$ one needs to include higher-order terms.

We extended this analysis to the full set of 15 order parameters by expanding around this solution to second order in $\phi_{A,B}$ and ${\vec S_A}$, ${\vec S_B}$. We found after long algebra  that (i) the minimum of $F^{(4)}$ is still at $-(7/3) |\beta| R^4$, and (ii)  the order parameter manifold at the minimum is parameterized in terms of Hopf coordinates and variables $\varepsilon$ and $r$ as
\begin{equation}
\begin{aligned}
&\vec S_{+} = r (0, 0, 1- \frac{\varepsilon^2}{2} \sin^2 \theta) \label{n_2}\\
&\vec S_A = r \varepsilon (\sin \theta \cos \psi_1, \sin \theta \sin \psi_1, \cos \theta \cos \psi_2); \\
&\vec S_B = r \varepsilon (\sin \theta \sin \psi_1, -\sin \theta \cos \psi_1, -\cos \theta \sin \psi_2); \\
&\phi= r(1-\frac{\varepsilon^2}{2});~ \phi_A = r\varepsilon \cos \theta \cos \psi_2;~ \phi_B = -r\varepsilon \cos \theta \sin \psi_2; \\
& \vec S_{-} = r (-\frac{\varepsilon^2}{2} \sin 2\theta \cos
\psi_+
, -\frac{\varepsilon^2}{2} \sin 2\theta \sin
\psi_+, 1 - \frac{\varepsilon^2}{2} \cos^2 \theta), \nonumber
\end{aligned}
\end{equation}
 where $\psi_+ = \psi_1 + \psi_2$, and we
  directed
  $\vec S_{+}$ along ${\hat z}$.  In terms of these variables, $R^2 = 3r^2 \left(1 + 3 \varepsilon^4/16 \left(1 - (1/9) \cos{4\theta}\right) + O(\varepsilon^6)\right)$.
The result $F^{(4)} =-(7/3) |\beta| R^4$ is also valid up to $O(\varepsilon^6)$.
Note that pure inter-valley order ($\phi=S_\pm=0$) is not a part of the order parameter manifold.
The seven independent variables in \eqref{n_2}, together with the requirement that $R^2$ is fixed, yield 6 Goldstone modes.
This can also be shown more rigorously by noticing that for $\beta<0$, SU(4) symmetry is broken down to SU(3)$\otimes$U(1) \cite{Shrock2010symmetry}.
The broken symmetry is described by
the coset  SU(4)/[SU(3)$\otimes$U(1)] with 15-8-1=6 generators, corresponding to the 6 Goldstone modes~\cite{SM}.

{\it{\bf Reconstructed fermionic dispersion.}}~~~Upon gap opening, the initial four-fold (spin and valley) degeneracy of the electronic dispersion in the 6- and 2-patch model is lifted. For the 6-patch model, we verified that the states remain doubly degenerate for any configuration from the order parameter manifold~\cite{SM}.
The easiest way to see this is to consider the state with $s^{+-}$ valley order: it splits chemical potentials in the two valleys but preserves spin degeneracy. For density-wave orders, the Fermi surfaces get reconstructed, but the bands remain two-fold degenerate. For the 2-patch model, the situation is different:  a four-fold degenerate Fermi level splits into a non-degenerate level and a 3-fold degenerate one, consistent with the residual SU(3)$\otimes$U(1) symmetry.  This again holds for any configuration from the order parameter manifold and can be seen most directly by restricting to intra-valley order with, e.g., $S_2=0$ and $S^{z}_1=\phi \neq 0$. This order shifts the non-degenerate level by $3\phi$ and the 3-fold degenerate level by $-\phi$. We show the energy splitting in Fig. \ref{band_split}.

{\it{\bf Conclusions. }}~~~
In this work we considered two itinerant patch models for TBG: the 2-patch model for small doping away from charge neutrality, when Fermi surfaces still form pockets near Dirac points $K$, $K'$, and the 6-patch model for Van Hove filling with 3 Van Hove points for each valley.
We analyzed potential instabilities in the particle-hole channel and derived the corresponding Landau functional. We argued that in both cases the largest and (almost) equal couplings are for a set of 15 order parameters. These 15 order parameters form an adjoint representation of SU(4), and the corresponding Landau functional is SU(4) symmetric.
We evaluated the prefactor $\beta$ for the fourth-order term in the Landau functional and found that it has opposite sign in the 6-
and 2-patch model.  In the 6-patch model, $\beta>0$. In this situation, the manifold of ordered states has 8 Goldstone modes, and the initially 4-fold degenerate energy level splits into 2 doubly degenerate levels. In the 2-patch model, $\beta<0$. Here we found a different order parameter manifold with 6 Goldstones and splitting of the 4-fold degenerate level into one non-degenerate level and 3 degenerate ones.
Our results describe the formation of a symmetry-broken ground state in TBG, either near Van Hove filling or for pocketed Fermi surfaces, which experimentally are near $n\approx 2$ or $n \approx 1$.
In both cases, we found intra-valley orders that go beyond a difference in the occupation of flavors, and inter-valley orders (inter-valley coherent states), which at Van Hove filling form real and imaginary density-waves.
%
%The itinerant approach is a good starting point to describe ordered states emerging out of a metal.
% For a system like TBG, it is interesting to connect to the opposite strong coupling limit, where the kinetic energy of fermions is considered to be a weak perturbation.
%Interestingly, the
The splitting of energy levels ( $4 \to 1+3$ and $4 \to 2+2$ ) and the number of Golstone modes (8 and 6) match some of the results for Chern insulators~\cite{Khalaf2020soft},
 although in our case the number of Goldstone modes
 is not directly related to the Chern numbers.
 %(a full agreement is possible if the topology of the full bands is incorporated into our approach).
Another similarity to strong-coupling approaches is the large number of degenerate ground states that we find due to the large symmetry. This has been discussed as a possible explanation for variations in experimental phase diagrams.

\acknowledgments{We thank R. Fernandes, L. Levitov, H. Polshyn,
 G. Tarnopolsky, O. Vafek, A. Vainshtein and A. Vishwanath for fruitful discussions. The work by D.V.C and A.V.C. was supported by U.S. Department of Energy, Office of Science, Basic Energy Sciences, under Award No. DE-SC0014402.
L.C. was supported by the U.S. Department of Energy (DOE), Of- fice of Basic Energy Sciences, under Contract No. DE- SC0012704. Y.W. was supported by NSF under award number DMR-2045871.
D.V.C. and A.V.C. also acknowledge the hospitality of KITP at Santa Barbara. The part of research done at KITP was supported in part by the National Science Foundation under Grant No. NSF PHY-1748958. D.V.C. gratefully acknowledges support from  Doctoral Dissertation  and Larkin Fellowships at the University of Minnesota.}
\bibliography{biblio}

\begin{widetext}
\section{Supplementary materials for: SU(4) symmetry in twisted bilayer graphene - an itinerant perspective}

\section{Order parameters and leading instabilities within the 6-patch model}

The order parameters in the 6-patch model have been analyzed in Ref.~\cite{Chichinadze2020magnet} and we just state the results here. The total number of bilinears  made of patch fermions is determined by the spin and valley composition, which combine to form generators of SU(4),  and the total number of patches within a given valley, which  is $N=3$ in our 6-patch model.
Assuming maximal possible symmetry,
this yields 143 orthonormal generators of an SU(4N)= SU(12), and 143 corresponding bilinears, which form the adjoint representation of SU(12). One can obtain the same number
by just counting the total number of bilinears with  $Q=0$ and different finite $Q$ between patches.  For the latter, an electron from each of 6 patches can transfer to 5 other patches, and the corresponding bilinear can be either 3-component spin $f^{\dagger} \vec \sigma f$, or single-component  density (charge) $f^{\dagger} f$, this gives 4 in total.
The total number of bilinears with a finite $Q$  is then $6\times 5 \times 4 = 120$.  For $Q=0$,  one can introduce a
  spin or charge order parameter  in each of the six patches, which brings the  total of  $Q=0$ orders to $6 \times 4 = 24$. However, the fully symmetric charge order parameter
     has to be discarded as it only renormalizes the chemical potential, so the actual number is 23.  Combining this with 120 bilinears with a finite $Q$, we end up with the same number of
     143.

To understand which bilinear order is more likely to develop upon lowering the temperature, we introduce infinitesimal
 orders $\Gamma^0_j(\vec Q)$ into the Hamiltonian and renormalize
 them
 in the ladder approximation. This leads to matrix equations for the dressed $\Gamma_j({\vec Q})$:
\beq
 \Gamma_j({\vec Q})=\Gamma_j^0({\vec Q}) +\Pi({\vec Q})\Lambda_{j, Q} \Gamma_j({\vec Q}),
 \label{ll}
  \eeq
  where $\Pi({\vec Q})$ is the polarization bubble at momentum ${\vec Q}$ and  $\Lambda_{j}$  (the matrices in patch space) contain the information about intra-patch and inter-patch interactions. The interactions with finite momentum transfer  depend on the relative strength of assisted hopping terms and Hubbard terms.
  We follow Ref. \cite{Kang2018strong}
  and use $\alpha_T$ as a parameter, which measures their relative strength.

Diagonalizing the equation, we obtain
\beq
{\bar \Gamma}_{j}({\vec Q}) = \frac{{\bar \Gamma}^0_{j}({\vec Q})}{1 - \Pi({\vec Q})\lambda_{j \, Q}},
\eeq
where ${\bar \Gamma}_{j}({\vec Q})$ are linear combinations of $\Gamma_{j}({\vec Q})$, and  $\lambda_{j \, Q}$ are the eigenvalues of the matrix equation (\ref{ll}).

The 143 bilinears can be combined into linear combinations
forming
one $s$ and two $d-$harmonics with different symmetry properties upon rotation by $\pi/3$.  There are 47 $s-$wave bilinears and $48+48$ $d-$wave bilinears. They decouple in the matrix equation. Calculations, presented in~\cite{Chichinadze2020magnet}, show that the couplings in $s-$wave channels are larger.  Out of 47 $s$-wave bilinears there are 7 with $Q=0$,  $16$ with momentum $Q_s$, $16$ with momentum $Q_m$, and $8$ with momentum $Q_l$, see  Fig. \ref{qvectors}.
  One can construct two different bilinears with each $Q_s$ and $Q_m$ (e.g., with $Q_s$ between patches labeled by $1$ and $2'$ and by $2$ and $1'$ in Fig. \ref{qvectors}), this further splits the bilinears with $Q_s$ and $Q_m$ into symmetric and anti-symmetric combinations with different $\lambda^+_Q$ and $\lambda^{-}_Q$.  All together, this creates 5 different sub-classes of bilinears with non-zero $Q$, with 8 elements in each sub-class.

 %%%%%%%%%%%%%%%%%%%%%%%%%%%%%%%%%%%%%%%%%%%%%%%%%%%%%%%%%%%%%%
\begin{figure}[h]
\includegraphics[width=0.25\linewidth]{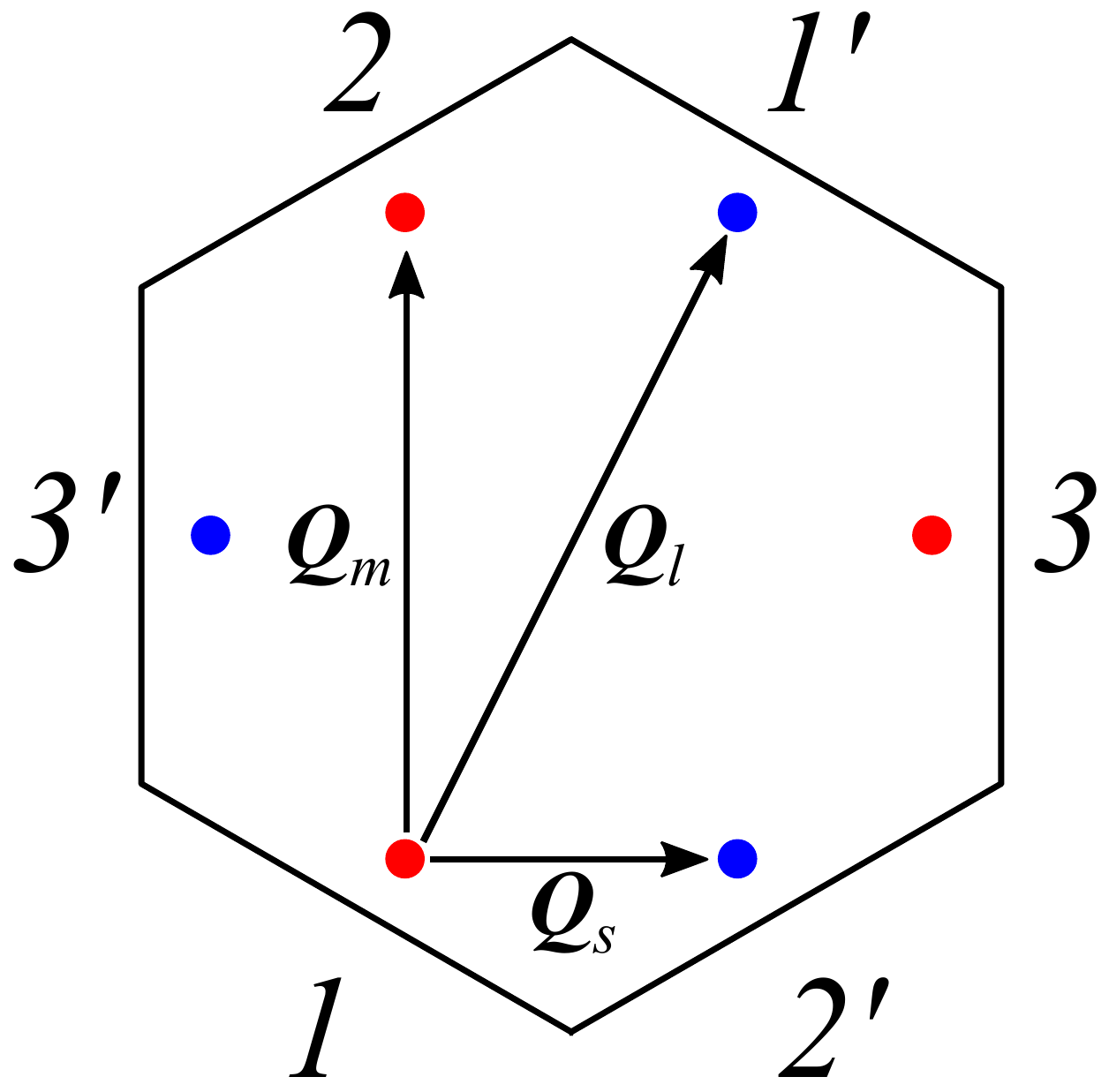}
\centering{}\caption{Three different vectors $Q_S$, $Q_m$, and $Q_l$ in the 6-patch model. }
\label{qvectors}
\end{figure}
%%%%%%%%%%%%%%%%%%%%%%%%%%%%%%%%%%%%%%%%%%%%%%%%%%%%%%%%%%%%%%

It turns out that the largest couplings are $\lambda_0$ ($Q=0$) and $\lambda^+_Q$ for $Q=Q_s$ These two couplings are positive (attractive),
substantially larger than other couplings, and
 very close to each other for a wide range of $\alpha_T$ (and are strictly degenerate for a particular value of $\alpha_T$).
We show these two largest couplings in Fig. \ref{lambda}.
Neglecting the
small
difference between these two $\lambda$'s
and polarization operators,
 we end up with the model of 7+8=15 identical  bilinears.  They can be viewed as forming the adjoint representation of SU(4).

 %%%%%%%%%%%%%%%%%%%%%%%%%%%%%%%%%%%%%%%%%%%%%%%%%%%%%%%%%%%%%%
\begin{figure}[h]
\includegraphics[width=0.5\linewidth]{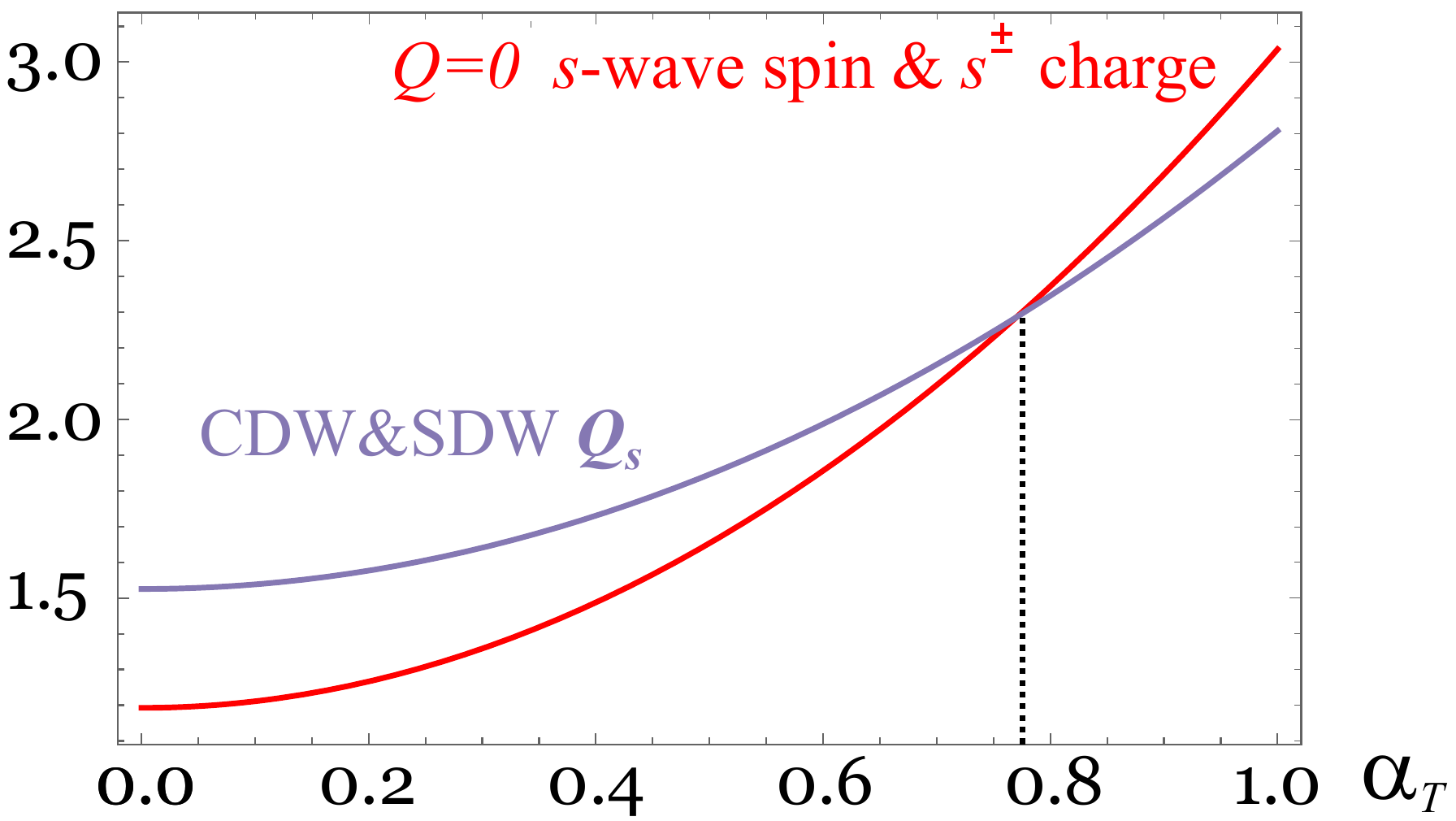}
\centering{}\caption{
The largest eigenvalues in the 6-patch model. Red line -- the eigenvalue for $Q=0$ orders (degenerate $s-$wave spin order and $s^{\pm}-$ charge (valley) order). Blue line - the eigenvalue for degenerate spin and charge density wave orders
 with  momentum transfer $Q_s$.}
\label{lambda}
\end{figure}
%%%%%%%%%%%%%%%%%%%%%%%%%%%%%%%%%%%%%%%%%%%%%%%%%%%%%%%%%%%%%%

\section{Order parameters and leading instabilities within the 2-patch model}

For smaller doping, the Fermi surfaces
are nearly circular, forming arcs near Dirac $K$  and $K'$ points, and filled states are located inside small pockets near these points.  It is essential that each pocket  now contains fermions from both valleys.
Strictly speaking,  Fermi energies of the two valleys merge only along the diagonals (and this holds also for larger fillings,  even near the Van Hove one (Ref. \cite{Wang2021}).  For small enough filling, however, the difference between the locations of the Fermi surfaces for the two valleys is small for all Fermi points.

We take
 the characteristic itinerant behavior near
the filling corresponding to $n=1$ as the one where, on one hand, Fermi surfaces for the two valleys are
 almost degenerate and on the other, the dispersion around the Fermi surface is
still approximately linear in momentum deviation from a Dirac point.
We then  associate pockets with patches and assume valley degeneracy within a patch.
This way we obtain an effective model with 2 patches, one at $K$, another at $K'$,  see Fig. \ref{2patchexpl}.
 We label patches $1$ and $2$ and introduce the valley index $v = \pm$.
There are six generally different couplings in the 2-patch model (see Fig. \ref{2patchexpl}). The
 interaction Hamiltonian reads:
\begin{equation}
\begin{gathered}
H^{Int}_{2p}=    \sum_{i=1}^{2} \sum_{v,v'=+,-}  \biggr[ u_0   f_{iv}^\dagger f_{iv} f_{iv}^\dagger f_{iv}  + v_0
f_{iv}^\dagger f_{iv} f_{iv'}^\dagger f_{iv'} + u_1 f_{iv}^\dagger f_{iv} f_{i+1v}^\dagger f_{i+1v} + v_1 f_{iv}^\dagger f_{iv} f_{i+1v'}^\dagger f_{i+1v'}
+ j  f_{iv}^\dagger f_{i+1v} f_{i+1v}^\dagger f_{iv}   + g  f_{iv}^\dagger f_{i+1v} f_{iv'}^\dagger f_{i+1v'}   \biggr].
\end{gathered}
\label{eq:iap2}
\end{equation}
 For Kang-Vafek model,
 %AC
  which we will use,
  $u_0 = u_1 = v_0 = v_1 =u$, and $j = g$.

 Fermionic bilinears for the 2-patch model
 can be cast into two  groups:
generalized Stoner (Pomeranchuk) instabilities (the one with $\vec Q =0$) and density-wave instabilities with momentum transfer $\vec Q$ between $K$ and $K'$.
 The same count as for the 6-patch model yields $15$ bilinears with $Q=0$ and $32$ with a finite $Q$, bringing the total to $63$.
 With maximal possible symmetry, they would
 form the adjoint representation of $SU(4*2) = SU(8)$.

We introduce the corresponding 64  (=63+1)  patch order parameters (the expectation values of fermionic bilinears) as
\begin{equation}
\begin{gathered}
\Delta_{1+}^c = \langle f_{1+}^{\dag} f_{1+} \rangle, \;  \Delta_{2+}^c = \langle f_{2+}^{\dag} f_{2+} \rangle, \; \Delta_{1-}^c = \langle f_{1-}^{\dag} f_{1-} \rangle, \; \Delta_{2-}^c = \langle f_{2-}^{\dag} f_{2-} \rangle, \\
\Delta_{1+}^s = \langle f_{1+}^{\dag} \vec \sigma f_{1+} \rangle, \;  \Delta_{2+}^s = \langle f_{2+}^{\dag} \vec \sigma f_{2+} \rangle, \; \Delta_{1-}^s = \langle f_{1-}^{\dag} \vec \sigma f_{1-} \rangle, \; \Delta_{2-}^s = \langle f_{2-}^{\dag} \vec \sigma f_{2-} \rangle, \\
\Delta_{1+-}^c = \langle f_{1+}^{\dag} f_{1-} \rangle, \;  \Delta_{2+-}^c = \langle f_{2+}^{\dag} f_{2-} \rangle, \; \Delta_{1-+}^c = \langle f_{1-}^{\dag} f_{1+} \rangle, \; \Delta_{2-+}^c = \langle f_{2-}^{\dag} f_{2+} \rangle, \\
\Delta_{1+-}^s = \langle f_{1+}^{\dag} \vec \sigma f_{1-} \rangle, \;  \Delta_{2+-}^s = \langle f_{2+}^{\dag} \vec \sigma f_{2-} \rangle, \; \Delta_{1-+}^s = \langle f_{1-}^{\dag} \vec \sigma f_{1+} \rangle, \; \Delta_{2-+}^s = \langle f_{2-}^{\dag} \vec \sigma f_{2+} \rangle, \\
\Delta_{12+}^c = \langle f_{1+}^{\dag} f_{2+} \rangle, \; \Delta_{12-}^c = \langle f_{1-}^{\dag} f_{2-} \rangle, \; \Delta_{12+}^s = \langle f_{1+}^{\dag} \vec \sigma f_{2+} \rangle, \; \Delta_{12-}^s = \langle f_{1-}^{\dag} \vec \sigma f_{2-} \rangle, \\
\Delta_{21+}^c = \langle f_{2+}^{\dag} f_{1+} \rangle, \; \Delta_{21-}^c = \langle f_{2-}^{\dag} f_{1-} \rangle, \; \Delta_{21+}^s = \langle f_{2+}^{\dag} \vec \sigma f_{1+} \rangle, \; \Delta_{21-}^s = \langle f_{2-}^{\dag} \vec \sigma f_{1-} \rangle, \\
\Delta_{12+-}^c = \langle f_{1+}^{\dag} f_{2-} \rangle, \; \Delta_{12-+}^c = \langle f_{1-}^{\dag} f_{2+} \rangle, \; \Delta_{12+-}^s = \langle f_{1+}^{\dag} \vec \sigma f_{2-} \rangle, \; \Delta_{12-+}^s = \langle f_{1-}^{\dag} \vec \sigma f_{2+} \rangle \\
\Delta_{21+-}^c = \langle f_{2+}^{\dag} f_{1-} \rangle, \; \Delta_{21-+}^c = \langle f_{2-}^{\dag} f_{1+} \rangle, \; \Delta_{21+-}^s = \langle f_{2+}^{\dag} \vec \sigma f_{1-} \rangle, \; \Delta_{21-+}^s = \langle f_{2-}^{\dag} \vec \sigma f_{1+} \rangle.
\end{gathered}
\end{equation}
We follow the standard procedure,
i.e., we
treat the bare order parameters $\Delta^{(0)}$  as infinitesimally small, add
$\Delta_{1+}^{c,(0)}  f_{1+}^{\dag} f_{1+}+ ...$  to the Hamiltonian, and renormalize $\Delta_{1+}^{c,(0)}$ and all other
$\Delta^{(0)}$  by the interaction.  Graphically, $\Delta^{(0)}$  are represented as  two-particle  vertices. It is convenient to combine vertices into combinations, separating spin and charge channels, $Q=0$ and a finite $Q$ and intra-valley and inter-valley order parameters, which we label below as intra and inter, respectively.  We have, skipping the index $(0)$ for simplicity,
\begin{equation}
\begin{gathered}
\Gamma_c^{intra} (0) = \left( \Delta_{1+}^c, \Delta_{2+}^c, \Delta_{1-}^c, \Delta_{2-}^c \right), \\
\Gamma_s^{intra} (0) = \left( \Delta_{1+}^s, \Delta_{2+}^s, \Delta_{1-}^s, \Delta_{2-}^s \right), \\
\Gamma_c^{inter} (0) = \left( \Delta_{1+-}^c, \Delta_{2+-}^c, \Delta_{1-+}^c, \Delta_{2-+}^c \right), \\
\Gamma_s^{inter} (0) = \left( \Delta_{1+-}^s, \Delta_{2+-}^s, \Delta_{1-+}^s, \Delta_{2-+}^s \right), \\
\Gamma_c^{intra} (Q) = \left( \Delta_{12+}^c, \Delta_{12-}^c, \Delta_{21+}^c, \Delta_{21-}^c \right), \\
\Gamma_s^{intra} (Q) = \left( \Delta_{12+}^s, \Delta_{12-}^s, \Delta_{21+}^s, \Delta_{21-}^s \right), \\
\Gamma_c^{inter} (Q) = \left( \Delta_{12+-}^c, \Delta_{12-+}^c, \Delta_{21+-}^c, \Delta_{21-+}^c \right), \\
\Gamma_s^{inter} (Q) = \left( \Delta_{12+-}^s, \Delta_{12-+}^s, \Delta_{21+-}^s, \Delta_{21-+}^s \right)
\end{gathered}
\end{equation}

The dressed vertices $\Gamma$ are expressed via $\Gamma^{(0)}$ by Eq. (\ref{ll}).
For intra-valley charge and spin Pomeranchuk orders the interaction matrices $\Lambda$ are
\begin{equation}
\Lambda^{C,intra}_{Pom} =
\begin{pmatrix}
-u & j-2u & -2u & -2u \\
j -2u & -u & -2u & -2u \\
-2u & -2u & -u & j-2u \\
-2u & -2u & j-2u & -u
\end{pmatrix}, \;\;\;
\Lambda^{S,intra}_{Pom} =
\begin{pmatrix}
u & j & 0 & 0 \\
j  & u & 0 & 0 \\
0 & 0 & u & j \\
0 & 0 & j & u
\end{pmatrix}.
\end{equation}
The corresponding eigenvalues are
\begin{equation}
\begin{gathered}
\lambda_{intra,CPom}^s = \Pi(0) \left( j - 7u \right), \\
\lambda_{intra,CPom}^d = \Pi(0) \left( u-j \right), \\
\lambda_{intra,CPom}^{s\pm} = \Pi(0) \left( u+j \right).
\end{gathered}
\label{intraPom2patch_c}
\end{equation}
 where $\Pi (0)$ is the polarization operator at zero momentum transfer.
For $\lambda_{intra,CPom}^{s}$ the eigenvector is $(1,1,1,1)$, i.e., the state
has
$s-$wave
symmetry,
 while for $\lambda_{intra,CPom}^{s\pm}$ the eigenvector is $(-1,-1,1,1)$, and  the state is $s^{\pm}-$wave with respect to valley,
 i.e. it has $s$-wave symmetry, but the order parameter has opposite sign on the two valleys
 (here and below we use small $s$   for $s$-wave symmetry and large $S$ for spin order).
 For $\lambda_{intra,cPom}^d$, there are two degenerate eigenvectors $(-1,1,0,0)$ and
$(0,0,-1,1)$. For spin intra-valley Pomeranchuk channel there are two eigenvalues
\begin{equation}
\begin{gathered}
\lambda_{intra,SPom}^s = \Pi(0) \left( u+j \right), \\
\lambda_{intra,SPom}^d = \Pi(0) \left( u-j \right).
\end{gathered}
\label{intraPom2patch_s}
\end{equation}
Each is doubly-degenerate with eigenvectors $(1,1,0,0), (0,0,1,1)$ and $(-1,1,0,0), (0,0,-1,1)$ correspondingly. Looking at Eqs. (\ref{intraPom2patch_c}) and (\ref{intraPom2patch_s}), we see that
  the leading intra-valley Pomeranchuk orders are  spin $s-$wave and charge $s^{\pm}$ orders.
For the $s^{\pm}$ order, $\Delta_{1+}^{c} = \Delta_{2+}^{c} = - \Delta_{1-}^{c} = - \Delta_{2-}^{c} = \phi$
( $=\Phi_{03}$ introduced in the main text). For the $s$-wave spin order  $\Delta_{1+}^{s} = \Delta_{2+}^{s} = \vec S_1 ;  \Delta_{1-}^{s} = \Delta_{2-}^{s} = \vec S_2$ ($= (\vec S_{+} \pm \vec S_{-})/\sqrt{2}
= (\Phi_{i,0} \pm \Phi_{i,3}) /\sqrt{2}$
 from the main text).

We now proceed to inter-valley Pomeranchuk orders.
There were no such order parameters in the 6-patch model,
because at the Van Hove filling the patches are valley-polarized. In the 2-patch model, these order parameters are present.
  The coupling matrices for charge and spin channels are identical because of the absence of valley mixing and are
\begin{equation}
\Lambda^{C,inter}_{Pom} = \Lambda^{S,inter}_{Pom} =
\begin{pmatrix}
u & j & 0 & 0 \\
j  & u & 0 & 0 \\
0 & 0 & u & j \\
0 & 0 & j & u
\end{pmatrix}.
\end{equation}
The eigenvectors and eigenvalues of this matrix are exactly the same as for intra-valley spin Pomeranchuk order parameter,
  i.e.,
 \begin{equation}
\begin{gathered}
\lambda_{inter,SPom}^s = \lambda_{inter,CPom}^s  = \Pi(0) \left( u+j \right), \\
\lambda_{inter,SPom}^d = \lambda_{inter,CPom}^d  = \Pi(0) \left( u-j \right).
\end{gathered}
\label{intraPom2patch_s1}
\end{equation}
   For positive $j$,
   the
   $s-$wave component has larger coupling than the $d-$wave component.  Keeping only orders with
    $\lambda_{inter,SPom}^s = \lambda_{inter,CPom}^s = \lambda_{intra,SPom}^s = \lambda_{intra,CPom}^{s+-}$, we
    end with
   15 degenerate inter-valley and intra-valley Pomeranchuk order parameters.

In simple words,  the mixed CDW/SDW state with $\vec{Q}_s$ in the 6-patch model at Van Hove doping is substituted by an inter-valley Pomeranchuk order,
which is now exactly degenerate with intra-valley Pomeranchuk order, not just at a special value of $\alpha_T$.
 The 15 Pomeranchuk order parameters as expressed via $\phi$ and ${\bf S}$ introduced in the main text %LC, as
 are
\begin{equation}
\begin{gathered}
\Delta_{1+-}^c = \Delta_{2+-}^c = \phi_A - i \phi_B, \; \Delta_{1-+}^c = \Delta_{2-+}^c = \phi_A + i \phi_B, \\
\Delta_{1+-}^s = \Delta_{2+-}^s = \vec S_A - i \vec S_B, \; \Delta_{1-+}^s = \Delta_{2-+}^s = \vec S_A + i \vec S_B.
\end{gathered}
\end{equation}
where, we remind,
\begin{equation}
\phi_A = \Phi_{0,1}, \; \phi_B = \Phi_{0,2}, \; \vec S_A = (\Phi_{1,1}, \Phi_{2,1}, \Phi_{3,1}), \; \vec S_B = (\Phi_{1,2}, \Phi_{2,2}, \Phi_{3,2}).
\end{equation}

We now proceed with density-wave orders. As in the case of Van Hove doping, there are inter-valley and intra-valley density wave order parameters.  In  the 2-patch model both are located at the same momentum  $Q$ between $K$ and $K'$.
  The matrices  $\Lambda$ for spin and charge intra-valley density wave orders are
\begin{equation}
\Lambda^{intra}_{CDW} = (u-2j) \mathbbm{1}_{4\times4}, \;\;\;
\Lambda^{intra}_{SDW} = u \mathbbm{1}_{4\times4},
\end{equation}
where $\mathbbm{1}_{4\times4}$ -- is a $4 \times 4$ unit matrix. The eigenvalues for spin channel are all equal $u$ and the eigenvalues for charge channel
are
all equal to $u-2j$. Then
\begin{equation}
\begin{gathered}
\lambda^{intra}_{CDW} = \Pi (\vec Q) (u-2j), \\
\lambda^{intra}_{SDW} = \Pi (\vec Q) u.
\end{gathered}
\end{equation}
The matrices $\Lambda$ for spin and charge inter-valley density wave orders are diagonal
\begin{equation}
\Lambda^{inter}_{CDW} = \Lambda^{inter}_{SDW} = u \mathbbm{1}_{4\times4}.
\end{equation}
 and $\lambda^{inter}_{CDW} = \lambda^{inter}_{SDW}=\Pi (\vec Q) u$.
 We found numerically that $\Pi (Q) \simeq \frac{2}{3} \Pi(0)$. Then $\lambda$'s  for the density-wave orders are smaller than for 15 Pomeranchuk orders with $\lambda = (u+j) \Pi (0)$, i.e., we can neglect density-wave order parameters.

%%%%%%%%%%%%%%%%%%%%%%%%%%%%%%%%%%%%%%%%%%%%%%%%%%%%%%%%%%%%%%
\begin{figure}[h]
\includegraphics[width=0.6\linewidth]{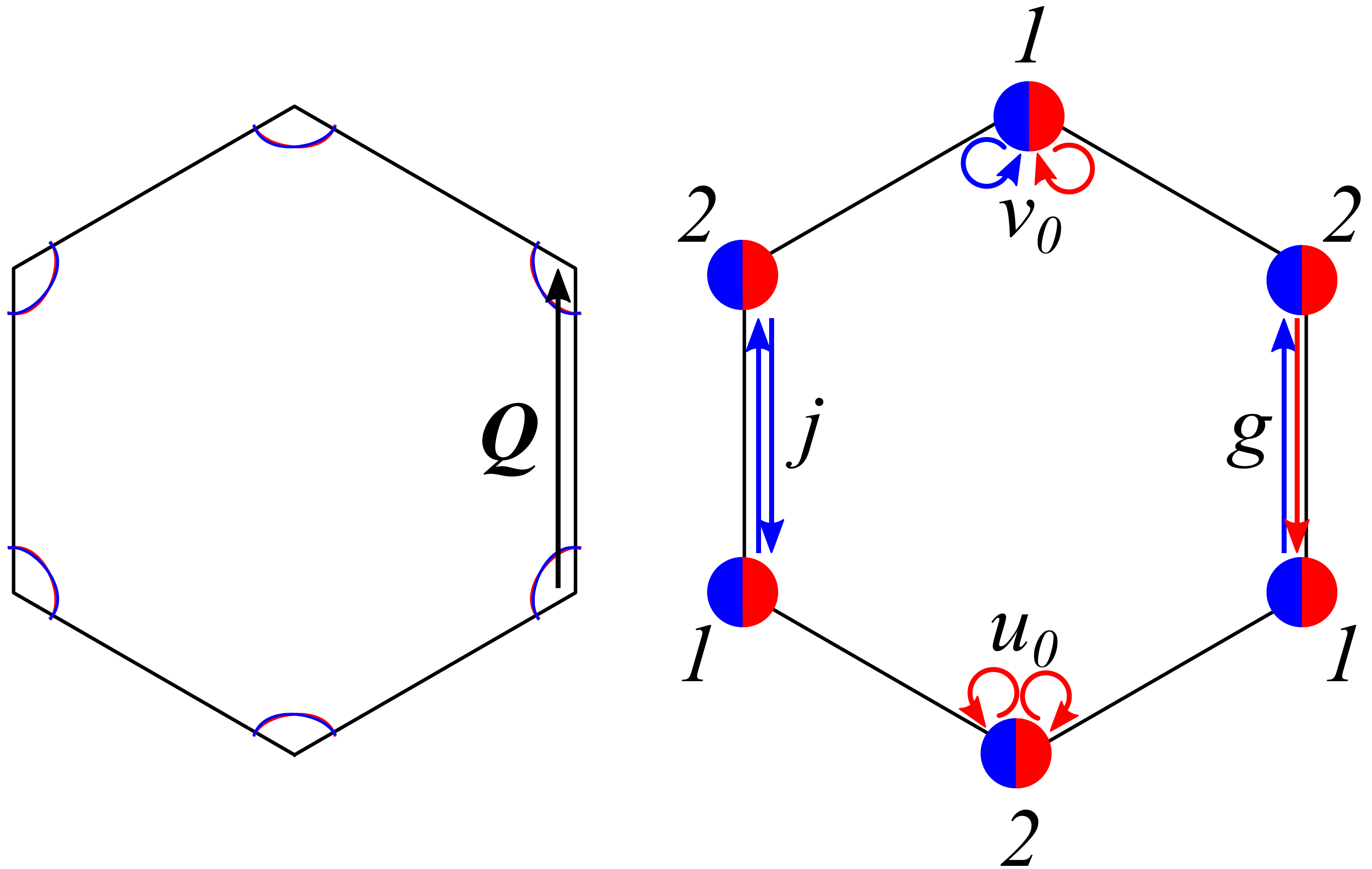}
\centering{}\caption{The two-patch model.
 The left panel shows Fermi pockets in the vicinity of $K$ and $K'$ points, with nearly degenerate dispersions for the two valleys. The right panel shows the interactions between patch electrons.}
\label{2patchexpl}
\end{figure}
%%%%%%%%%%%%%%%%%%%%%%%%%%%%%%%%%%%%%%%%%%%%%%%%%%%%%%%%%%%%%%

\section{Hubbard-Stratonovich transformation and the derivation of the Landau free energy}

The derivation of the Landau functional for the 6-patch model has been discussed in Ref. \cite{Chichinadze2020magnet}.
Here, we provide some details of derivation of the Landau functional
for the 2-patch model.  One can  straightforwardly verify
  that the free energy is the sum of
  two
   identical pieces for each of the 2 patches. We therefore focus on a single patch.

To perform the Hubbard-Stratonivich  transformation,  we introduce matrices of the Green's function ($\hat G_0$) and Hubbard-Stratonovich
fields for 15 Pomeranchuk orders
$\Phi_{i,j}$,  which couple to the fermions via the generators $\sigma_i \otimes \tau_j$.
 The  effective Hamiltonian of the
system is
\begin{equation}
H = \psi^{\dagger} G^{-1}\psi,
\end{equation}
where $\psi$ is the fermionic spinor and
\begin{equation}
G^{-1} = G_0^{-1} + \sum_{\{i,j\}} \Phi_{i,j}\sigma_i\otimes\tau_j.
\end{equation}
The sum runs over $i,j=0\ldots 3$ with $i=j=0$ excluded.
Integrating out fermions,  we
obtain
\begin{equation}
\mathrm{Tr} \; \mathrm{ln} \hat{G}^{-1}
=  \mathrm{Tr} \; \mathrm{ln} \left[ \hat{G}_{0}^{-1} (1 + \hat{G}_0 {\sum_{\{i,j\}}} \Phi_{i,j}\sigma_i\otimes\tau_j \right] = \mathrm{const.} + \mathrm{Tr} \; \mathrm{ln} \left[ 1 + \hat{G}_0 {\sum_{\{i,j\}}} \Phi_{i,j}\sigma_i\otimes\tau_j \right],
\end{equation}
where the trace is taken over spin
and valley
indices. Expanding the log in
$\Phi_{i,j}$, we obtain the
 Landau free energy. At quadratic order,
\begin{equation}
F^{(2)} =
\mathrm{Tr} [(\hat{G}_0 {\sum_{\{i,j\}}} \Phi_{i,j}\sigma_i\otimes\tau_j )^2] = \Pi (0) \left( \phi^2 + \phi_A^2 + \phi_B^2 + S_+^2 + S^2_- + S_A^2 + S_B^2 \right)\,,
\end{equation}
where we have used that $\mathrm{Tr}[\sigma_i\otimes\tau_j,\sigma_{i'}\otimes\tau_{j'}=4\delta_{i,i'}\delta_{j,j'}]$.
 This is the first term in Eq. (2) in the main text.
 Expanding  to cubic and quartic orders, we obtain the other two terms in Eq. (2) in the form, presented in Eqs. (3) and (4) of the main text.

\section{Calculation of the prefactor $\beta$ for the  quartic term in Landau free energy}

In this section we derive the prefactor $\beta$ for the quartic term in the Landau free energy. We do this for the Van Hove doping,
and
for a pocketed  Fermi surface.
In the latter case, we calculate $\beta$ with two models for the Fermi pockets, a finite offset from Van Hove doping and a linear dispersion, and show that the qualitative behavior does not depend on these details.

We start with box diagrams for the orders with $Q=0$.  For the 2-patch model these are the only terms which we need, once we neglect the difference between the dispersions of fermions from different valleys near $K$.
 For the 6-patch model, the prefactors for the terms with the fourth power of the orders with a finite $Q$ and for cross-terms with the products of squares of orders with $Q=0$ and with finite $Q$ are not the same, but we argue that for the dispersion appropriate for TBG, the difference is relatively small.

%%%%%%%%%%%%%%%%%%%%%%%%%%%%%%%%%%%%%%%%%%%%%%%%%%%%%%%%%%%%%%
\begin{figure}[h]
\includegraphics[width=0.6\linewidth]{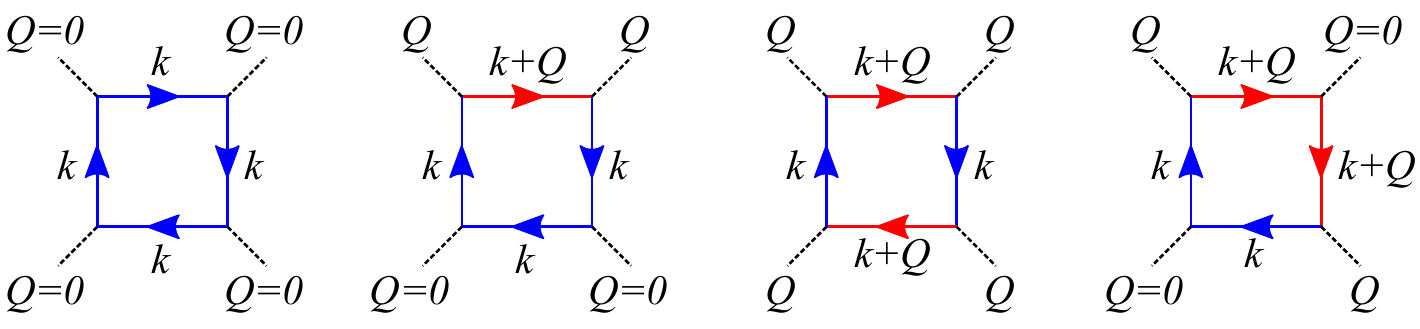}
\centering{}\caption{Box diagrams for prefactors $\beta_{1,2,3}$ in the six-patch model.
  Red and blue arrows depict fermionic Green's functions of electrons from different valleys.}
\label{box_diag}
\end{figure}
%%%%%%%%%%%%%%%%%%%%%%%%%%%%%%%%%%%%%%%%%%%%%%%%%%%%%%%%%%%%%%

\subsection{$\beta$ at the Van Hove doping}

The prefactor $\beta$ for $Q=0$ orders  is expressed via the integral over momentum and the sum over  fermionic Matsubara frequency of the product of four fermionic propagators with the same frequency and momentum:
\begin{equation}
\beta = 2 T \sum_{\omega_n} \int \frac{d^2 k}{4\pi^2} \left( \frac{1}{i \omega_n - \varepsilon} \right)^4
\label{ss}
\end{equation}

We compute $\beta$ using two approaches. In the first, we integrate over frequency first and then over momentum.
 This approach requires one to set a finite cutoff of momentum integration, but the final result does not depend on the cutoff.  In the second approach, we integrate over momentum first and then over frequency.  This way, we will not need
  to impose the cutoff on momentum integration.   We  obtain the same result for $\beta$ in both approaches.

\subsubsection{First approach}

Performing the sum over the fermionic Matsubara frequencies, we obtain from (\ref{ss}):
\begin{equation}
\beta = \frac{1}{96 \pi^2 T^3} \int d^2 k \frac{2-\cosh \frac{\varepsilon}{T}}{\cosh^4 \frac{\varepsilon}{2T}}
\label{ss_1}
\end{equation}
Without loss of generality, we choose the Van Hove point, around which the dispersion is $\epsilon_k = a k^2_k - b k^2_y$, where $a$ and $b$ are positive.  Rescaling the momenta to $k_x \sqrt{a} \rightarrow k_x$ and $k_y \sqrt{b} \rightarrow k_y$ and introducing  polar coordinates, we express $\beta$ as
\begin{equation}
\beta = \frac{1}{48 \pi^2 \sqrt{ab} T^2} \int_{0}^{\pi/2} \frac{d \theta}{\cos \theta} \int_0^{\Lambda \cos{\theta}/T} dx  \frac{2 - \cosh{x}}{\cosh^4 \dfrac{x}{2}},
\end{equation}
where $\Lambda$ is the cutoff on $k^2$ and
Evaluating the integral over $x$,  we obtain
\begin{equation}
\beta = \frac{1}{3 \pi^2\sqrt{ab} T^2} \int_{0}^{\pi/2} \frac{d \theta}{\cos \theta} \frac{ \sinh^4 \frac{\Lambda \cos \theta}{2T}}{\sinh^3 \frac{\Lambda \cos \theta}{T}}.
\end{equation}
The integrand is singular at $\theta$ near $\pi/2$.  Expanding near $\pi/2$ as $\theta = \pi/2-\delta$ and introducing new  variable $y = \delta \Lambda/T$, we find
\begin{equation}
\beta = \frac{2}{3 \pi^2 \sqrt{ab} T^2} \int_{0}^{\infty} \frac{dy}{y} \frac{\sinh^4{\frac{y}{2}}}{\sinh^3 y}
 = \frac{1}{3 \pi^2 \sqrt{ab} T^2} \left(- \frac{7}{2} \zeta' (-2) \right) = \frac{0.0036}{\sqrt{ab} T^2}
\label{beta_assympt}
\end{equation}

In Fig. \ref{beta_num_and_app} we compare Eq. (\ref{beta_assympt}) with the result of full numerical evaluation of Eq. (\ref{ss_1}) for $\beta$.  We see that the agreement is very good.

%%%%%%%%%%%%%%%%%%%%%%%%%%%%%%%%%%%%%%%%%%%%%%%%%%%%%%%%%%%%%%
\begin{figure}[h]
\includegraphics[width=0.6\linewidth]{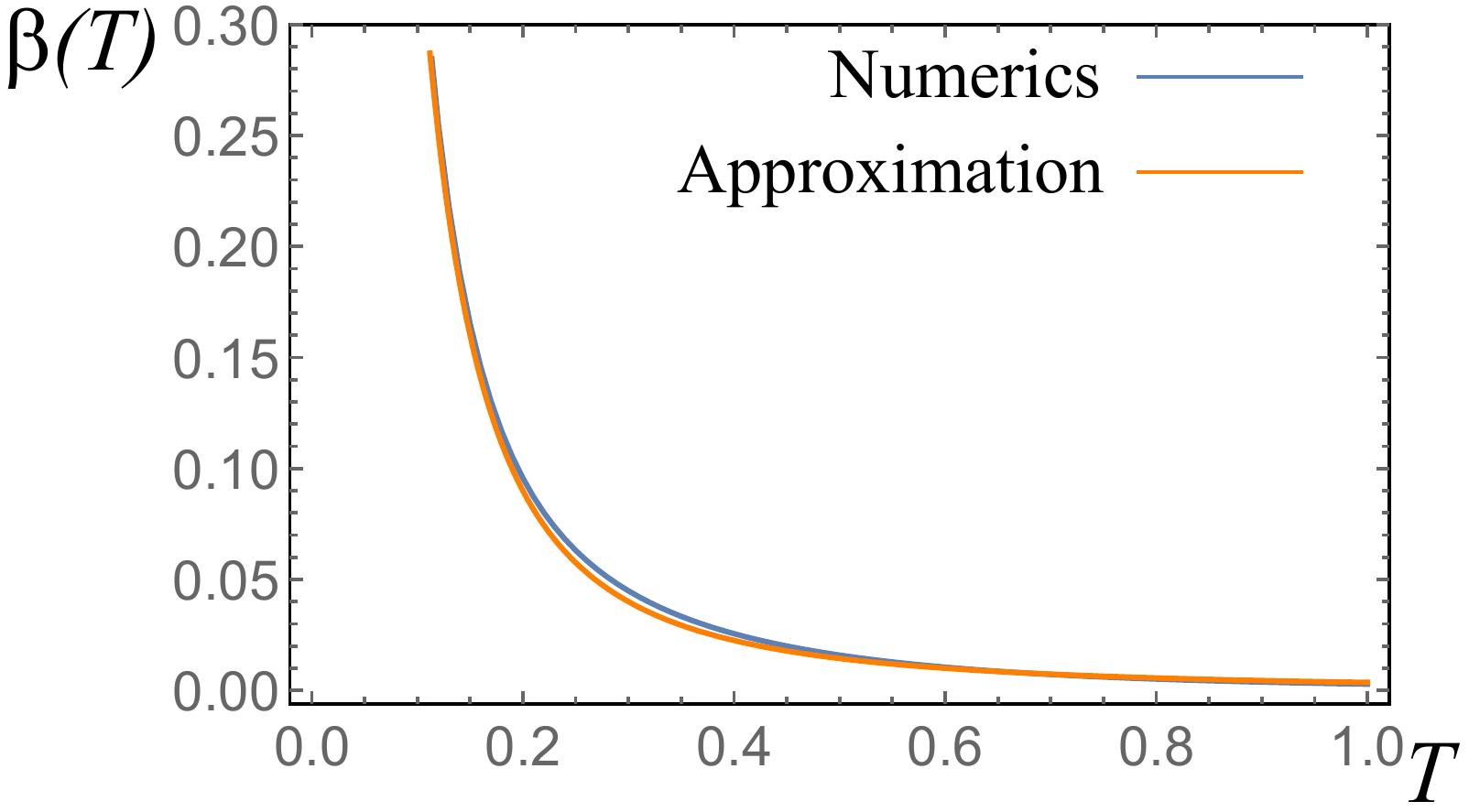}
\centering{}\caption{
$\beta(T)$ as a function of $T$. The analytical formula, Eq. (\ref{beta_assympt}) matches the  result of the full
 numerical integration almost perfectly. $\beta$ is in units of $1/(ab)^{3/2}$, and temperature is in units of $(ab)^{1/2}$, where $a$ and $b$ are the factors in the dispersion $\epsilon = ax^2-by^2$ }
\label{beta_num_and_app}
\end{figure}
%%%%%%%%%%%%%%%%%%%%%%%%%%%%%%%%%%%%%%%%%%%%%%%%%%%%%%%%%%%%%%

\subsubsection{Second approach}

We re-express Eq. (\ref{ss}) as
\be
\beta = T \sum_{\omega_m >0} \frac{1}{\omega^3_m} Y
\ee
where
\be
 Y = \frac{1}{2\pi^2 \sqrt{ab}} \int_0^\infty dz \int_0^\pi d \theta \left(\frac{1}{(z \cos{\theta} -i)^4} +  \frac{1}{(z \cos{\theta} +i)^4}\right)
\ee
Evaluating the integral over $\theta$ by taking the third derivative over $a =1/z$ of $\int_0^\pi d \theta/(\cos{\theta} -ia) = i\pi/\sqrt{a^2+1}$,  we obtain
\be
Y = \frac{1}{2\pi \sqrt{ab}} \int_0^\infty \frac{dz (2-3 z^2)}{(z+1)^{7/2}} = \frac{1}{3\pi \sqrt{ab}}
\ee
Using then
\be
T\sum_{\omega_m >0} \frac{1}{\omega^3_m} = \frac{1}{\pi^3 T^2} \sum_{n=0}^\infty \frac{1}{(2n+1)^3} = \frac{1}{\pi^3 T^2} \frac{7 \zeta[3]}{8}
\ee
we obtain
\be
\beta = \frac{1}{T^2 \sqrt{ab}} \frac{7 \zeta[3]}{24 \pi^4} = \frac{0.0036}{T^2 \sqrt{ab}}
\ee
 which is the same expression as in (\ref{beta_assympt}).

\subsection{$\beta$ away from the Van Hove doping}

\subsubsection{First approach}

 We next evaluate $\beta$ for the same dispersion, but for a finite  offset $\delta \mu$ from the Van Hove doping: $\varepsilon = ak_x^2 - bk_y^2 + \delta \mu$. The relevant variable in this case is $y = \delta \mu/T$.
Evaluating $\beta$ by integrating first over frequency, we obtain
\be
\beta = \frac{1}{192\pi^2 \sqrt{ab} T^2} I (y)
\ee
 where
 \be
I(y) = \int_{0}^{2\pi} \frac{d \theta}{\cos \theta} \int_{y}^{\Lambda \cos{\theta}/T +y}
 \frac{d x (2 - \cosh x)}{\cosh^4 \frac{x}{2}}.
\ee
Expanding again near $\theta = \pi/2$ and $3\pi /2$,
%(positive and negative $\cos \theta$ should be treated separately),
we obtain
\begin{equation}
I(y) =
%64
 32
\int_{0}^{\infty} \frac{dx}{x} \left( \frac{\sinh^4 \frac{x+y}{2}}{\sinh^3 (x+y)}
%-
+ \frac{\sinh^4 \frac{x-y}{2}}{\sinh^3 (x-y)} \right).
\end{equation}
The result of numerical evaluation of $I(y)$ is shown in Fig. \ref{Ivsy}. As one can see, $I(y)$ changes sign at $y \approx 2$, i.e., at  $T \sim \delta \mu$. For larger $y$, i.e., smaller $T$,  $\beta$  becomes negative.
At larger $\delta \mu$, i.e., at larger deviations from Van Hove density towards smaller doping, $\beta$ becomes
 negative  starting from progressively larger temperatures.
 %The value of $\beta$ decreases as the dispersion comes closer to a parabola.
 % For a purely parabolic dispersion $\epsilon = k^2/(2m) -\mu$, $\beta$ vanishes up to exponentially small terms $e^{-\mu/T}$. A negative $\beta$ then obviously comes from deviations from a parabolic dispersion.
%%%%%%%%%%%%%%%%%%%%%%%%%%%%%%%%%%%%%%%%%%%%%%%%%%%%%%%%%%%%%%
\begin{figure}[h]
\includegraphics[width=0.6\linewidth]{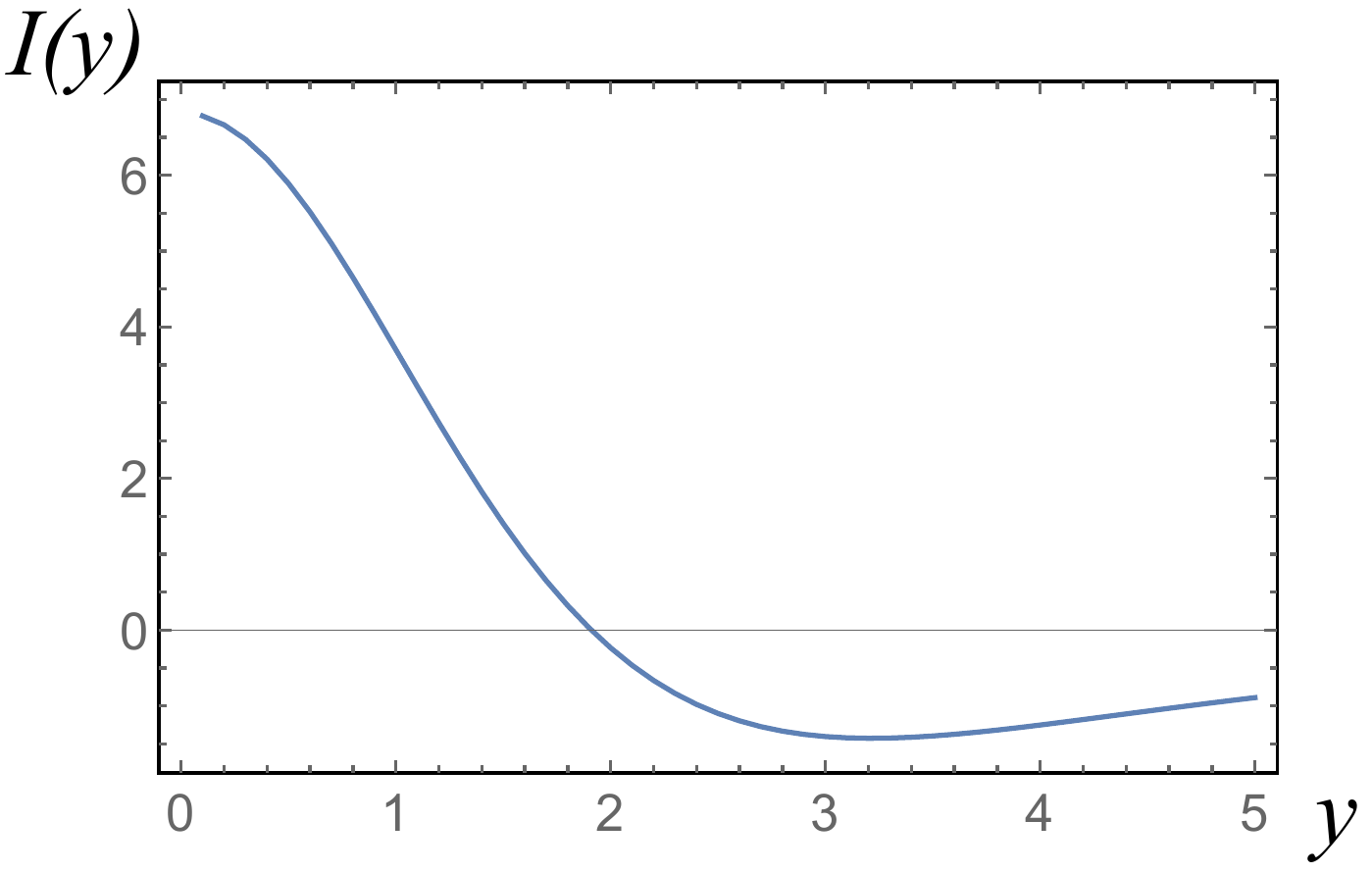}
\centering{}\caption{
$I(y)$ as a function of $y = \delta \mu/T$.
%One can see that the sign
 It changes sign  at $y\simeq 2$, i.e., at
 % which corresponds to
  $T \sim \delta \mu$.}
\label{Ivsy}
\end{figure}
%%%%%%%%%%%%%%%%%%%%%%%%%%%%%%%%%%%%%%%%%%%%%%%%%%%%%%%%%%%%%%

\subsubsection{Second approach}
We start again from
\be
\beta =  \frac{1}{96 \pi^2 T^3} \int d^2 k \frac{2-\cosh \frac{\varepsilon}{T}}{\cosh^4 \frac{\varepsilon}{2T}},
\ee
but calculate $\beta$ with a linear dispersion  $\epsilon=v |\vec k|-\mu$. We obtain
\begin{align}
\beta&=  \frac{1}{48 \pi T^2 v^2} \int_{-\mu/T}^{\Lambda/T} dx (T x +\mu)  \frac{2-\cosh x}{\cosh^4 \frac{x}{2}}\notag \\
&=\frac{1}{48 \pi T^2 v^2} \left[\frac{2}{\cosh^2\frac{\Lambda}{2T}}\left(T+(\mu+\Lambda)\tanh\frac{\Lambda}{2T} \right)-\frac{2T}{\cosh^2\frac{\mu}{2T}}\right]\,
\end{align}
where $\Lambda$ is again the UV cutoff.
For reasonable $\mu \leq \mathcal{O}(\Lambda)$, we again obtain $\beta<0$ at low temperatures. Furthermore, for $\Lambda \to \infty$, $\beta$ is negative for all $T$.
%%%%%%%%%%%%%%%%%%%%%%%%%%%%%%%%%%%%%%%%%%%%%%%%%%%%%%%%%%%%%%
\begin{figure}[h]
\includegraphics[width=0.4\linewidth]{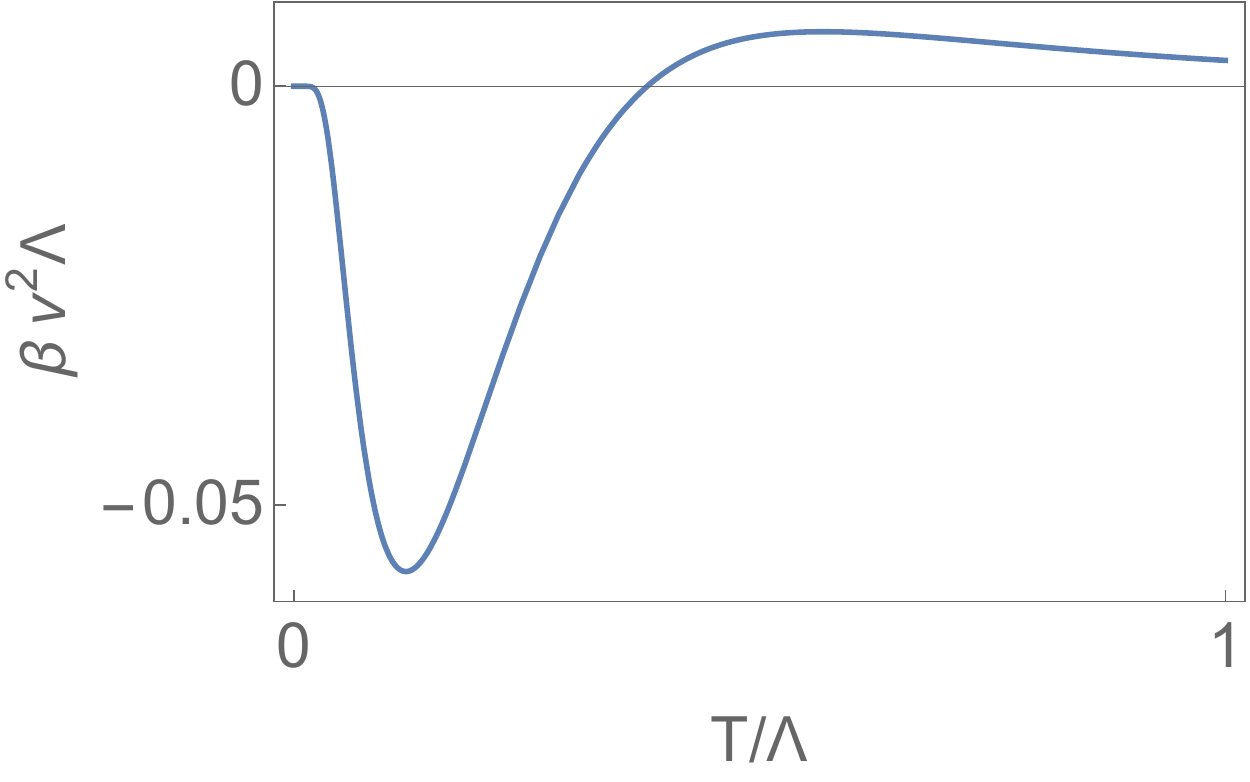}
\centering{}\caption{
$\beta$ as a function of temperature, computed using the linear dispersion $\epsilon=v|\vec k|-\mu$ with $\mu=0.2\Lambda$. }
\label{Ivsy_2}
\end{figure}
%%%%%%%%%%%%%%%%%%%%%%%%%%%%%%%%%%%%%%%%%%%%%%%%%%%%%%%%%%%%%%
%AC
 We also computed $\beta$ for the 2-patch model, using lattice dispersion for the two valleys.
  For $n$ near 1, we obtained $\beta (T)$ similar to the ones in Figs. \ref{Ivsy} and \ref{Ivsy_2}.

\subsection{$\beta_i$ for $Q=0$ and finite $Q$ orders at van Hove doping, 6-patch model}

As we said in the main text,  SU(4) symmetry in the 6-patch model is only approximate because the box diagrams for
 the orders with $Q=0$ and finite $Q$ contain different combinations of Green's functions.
   Specifically,
   there are
    three are different  prefactors $\beta_i$, which are given by the following integrals
$$
\beta_1 = \int G^4_k, \; \beta_2 = \int G^3_{k} G_{k+Q}, \; \beta_3 = \int G_{k}^2 G_{k+Q}^2.
$$
We now calculate these prefactors and check if they are close enough to be treated as equal.

We start by noticing that for the approximate SU(4) description of the 6-patch model, $Q$ connects neighboring van Hove points (belonging to different valleys). For definiteness, let's pick the two ``upper'' Van Hove points in the Brillouin zone (Fig. 1 in the main text).  The dispersion at these van Hove points can be generally expressed as
\begin{equation}
\varepsilon_1 = a_1 k_x^2 + c k_x k_y + b_1 k_y^2, \; \varepsilon_2 = a_1 k_x^2 - c k_x k_y + b_1 k_y^2.
\end{equation}
  We can rescale the momentum and incorporate $a_1$ into $x^2$ and $b_1$ into $y^2$. We label new momenta as ${\tilde k}$.
   The dispersions become
\begin{equation}
\varepsilon_1 =  \tilde{k}_x^2 + C \tilde{k}_x \tilde{k}_y +  \tilde{k}_y^2, \; \varepsilon_2 =  \tilde{k}_x^2 - C \tilde{k}_x \tilde{k}_y + \tilde{k}_y^2,
\end{equation}
where $C = \frac{c}{\sqrt{a_1 b_1}}$. The differential $d^2 k$ in the new rescaled variables reads $\frac{d^2 \tilde{k}}{\sqrt{a_1 b_1}}$. In polar coordinates,
\begin{equation}
\varepsilon_{1,2} = \tilde{k}^2 \pm C \tilde{k}^2 \sin \theta \cos \theta.
\end{equation}
The parameter $C$ depends on the underlying microscopic dispersion and has to be larger than $2$.

 The integral for $\beta_1$ can then be rewritten as
\begin{equation}
\beta_1 = 2T \sum_{\omega_n >0} \int \frac{d^2 k}{4\pi^2} \left(\frac{1}{(i \omega_n - \varepsilon)^4} +
\frac{1}{(-i \omega_n - \varepsilon)^4}\right)  =
T \sum_{\omega_n >0} \frac{1}{8\pi^2\sqrt{a_1 b_1}} \frac{1}{\omega_n^3} \int dz d\theta \left( \frac{1}{(i - z - C z \sin \theta \cos \theta)^4} + \frac{1}{(i + z + C z \sin \theta \cos \theta)^4} \right).
\end{equation}
A similar procedure can be done for $\beta_2$ and $\beta_3$
\begin{equation}
\begin{gathered}
\beta_2 = 2T \sum_{\omega_n >0} \frac{1}{8\pi^2\sqrt{a_1 b_1}} \frac{1}{\omega_n^3} \int dz d\theta \left( \frac{1}{(i - z - C z \sin \theta \cos \theta)^3 (i - z + C z \sin \theta \cos \theta)} + \frac{1}{(i + z + C z \sin \theta \cos \theta)^3 (i + z - C z \sin \theta \cos \theta)} \right), \\
\beta_3 = 2T \sum_{\omega_n >0} \frac{1}{8\pi^2\sqrt{a_1 b_1}} \frac{1}{\omega_n^3} \int dz d\theta \left( \frac{1}{(i - z - C z \sin \theta \cos \theta)^2 (i - z + C z \sin \theta \cos \theta)^2} + \frac{1}{(i + z + C z \sin \theta \cos \theta)^2 (i + z - C z \sin \theta \cos \theta)^2} \right).
\end{gathered}
\end{equation}
One can easily verify that for a generic $C >2$, the  integrals over $d\theta d z$ are convergent. This is different from the case of
a
nested Fermi surface and Van Hove points at the zone boundary \cite{NandkishorePRL2012}.
In
our
situation, $\beta_i = 2T \sum_{\omega_n >0} \frac{1}{8\pi^2\sqrt{a_1 b_1}} \frac{1}{\omega_n^3} {\tilde \beta}_i =
 (7 \zeta[3]/(64 \pi^5 T^2 \sqrt{a_1 b_1})) {\tilde \beta}_i$ are all proportional to $1/T^2$, and the ratios of different $\beta_i$ are the same as the ratios of different ${\tilde \beta}_i$.

 The results of numerical integration are shown in Fig. \ref{betas_plot}.
  For the microscopic dispersion which we and others used \cite{Yuan2018,Koshino2018PRX,Chichinadze2020magnet}
  $C$ is in the range from $\sim 3.5$ to $\sim 4.5$.
 We see that in this range $\beta_i$  differ by only 10-15 \%.  This justifies our approximation in which we treat $\beta_i$ as equal.

%%%%%%%%%%%%%%%%%%%%%%%%%%%%%%%%%%%%%%%%%%%%%%%%%%%%%%%%%%%%%%
\begin{figure}[h]
\includegraphics[width=0.6\linewidth]{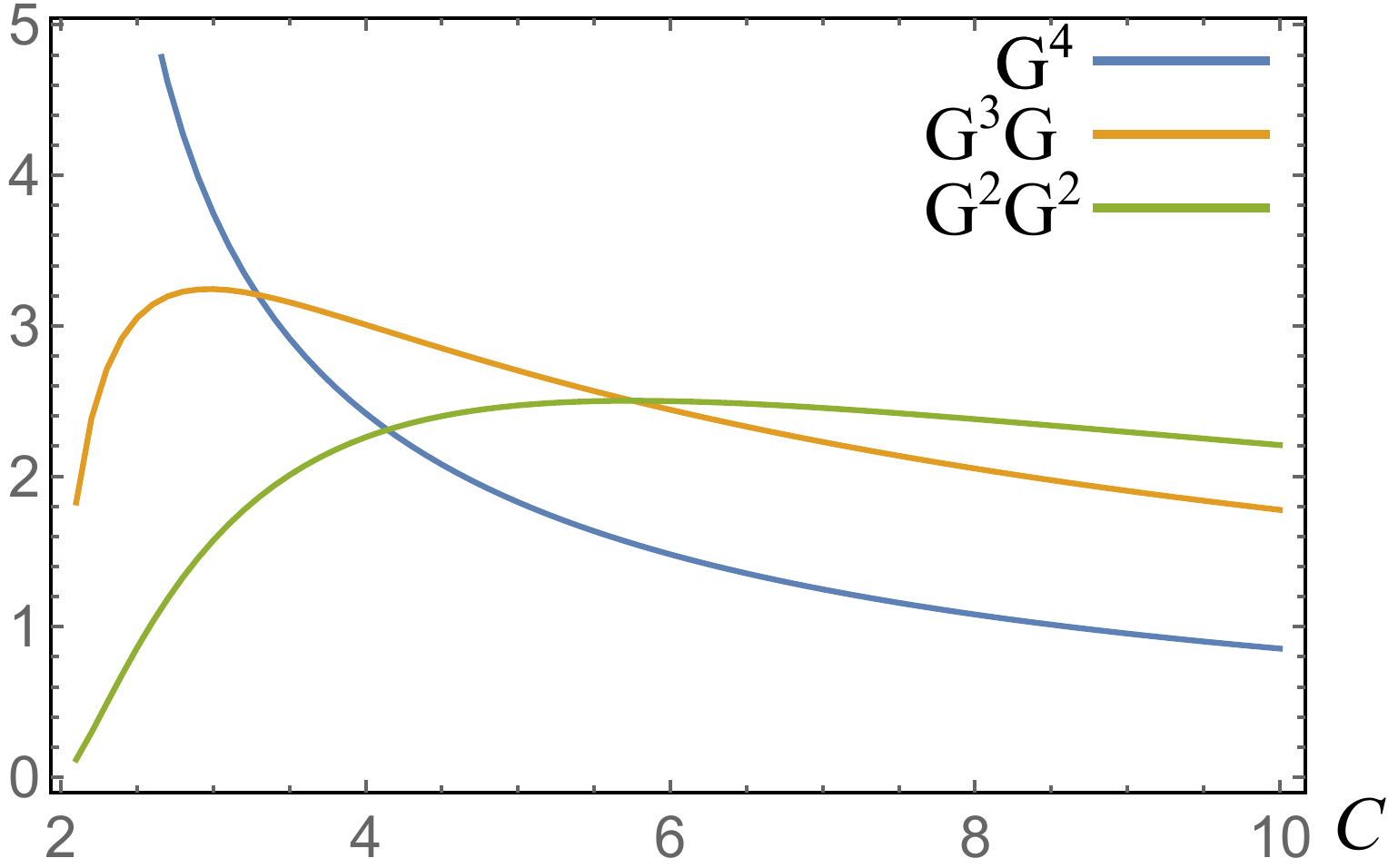}
\centering{}\caption{The results for ${\tilde \beta}$, up to common overall  factor.
  Blue line - ${\tilde \beta}_1$,  orange - ${\tilde \beta}_2$,  and green -- ${\tilde \beta}_3$.
}
\label{betas_plot}
\end{figure}
%%%%%%%%%%%%%%%%%%%%%%%%%%%%%%%%%%%%%%%%%%%%%%%%%%%%%%%%%%%%%%

\section{Minimization of the Landau free energy }

\subsection{Minimization  for $\beta>0$}

 For positive $\beta$,  $F^{(4)}$ contains $\beta R^4$ and the sum of full squares (the $C$ term
 in the main text)
  with the positive coefficient $4\beta$.
  The Landau functional $F$ then is at a minimum when the order parameters satisfy
\begin{equation}
\begin{gathered}
\left( \vec S_{+} \cdot \vec S_{-} \right) =\left( \vec S_A \cdot \vec S_{+} \right)=
 \left( \vec S_B \cdot \vec S_{+} \right)= \left(  \vec S_A \phi_A + \vec S_B \phi_B + \vec S_{-} \phi \right)=0, \\
 \left( \vec S_B \times \vec S_A + \phi \vec S_{+} \right)= \left( \vec S_A \times \vec S_{-} + \phi_B \vec S_{+} \right)=
  \left( \vec S_{-} \times \vec S_{B} + \phi_A \vec S_{+} \right) =0.
 \end{gathered}
 \label{min_cond}
\end{equation}
 For any configuration  that satisfies (\ref{min_cond}), $F= \alpha R^2 + \beta R^4$, and minimizing at $T < T_{ph}$, where $\alpha <0$,  we obtain the non-zero value of the total $R^2 =|\alpha|/(2\beta)$.  The conditions (\ref{min_cond}) are satisfied if we choose, e.g., ${\vec S}_+$ to be along $\hat{z}$:  ${\vec S}_+ = (0,0, S_+)$, and set other order parameters to be
\begin{equation}
\begin{gathered}
\vec S_{-} = (S_{x}, S_{y}, 0); \vec S_A = (A_x, A_y, 0); \vec S_B = (B_x, B_y, 0); \\
A_y B_x -A_x B_y  = -\phi S_{+}; A_x S_{y} -A_y S_{x}  = -\phi_B S_{+}; B_y S_{x}- B_x S_{y}= -\phi_A S_{+}.
\end{gathered}
\label{n_1}
\end{equation}
  An example
  of a configuration that satisfies (\ref{n_1}) is the one with  only intra-valley components  $\phi$ and ${\vec S}_{+-}$. Eq. (\ref{n_1}) allows
  two solutions: (i) $\phi \neq 0, {\vec S}_{\pm} =0$ and (ii) $\phi = 0, {\vec S}^2_{+} + {\vec S}^2_{-}= 2S^2$ and $\vec S_{+} \cdot \vec S_{-}=0$.  The first one describes the $s^{+-}$  valley
   order, the second  describes
   intra-valley ferromagnetism with  equal magnitudes of ${\bf S}_{1,2}$  in the two valleys, ${\vec S}^2_1= {\vec S}^2_2 = S^2$,  and arbitrary angle between  ${\vec S}_1$ and ${\vec S}_2$.

\subsection{Minimization for $\beta<0$. Intra-valley orders only}

Consider the case when $\beta <0$ and only intra-valley order parameters $\phi$, $S_{1}$ and $S_{2}$ are present. In this situation,
\begin{equation}
F = \alpha \left( \phi^2 + S_1^2 + S_2^2 \right) +  \beta \left( \phi^4 + 2 S_1^4 + 2 S_2^4 + 6 \phi^2 S_1^2 + 6 \phi^2 S_2^2 \right).
\label{Fintratrans}
\end{equation}
Three order parameters magnitudes can be now represented as components of a three-dimensional vector $(\phi, S_1, S_2)$ which we parameterize as
\begin{equation}
\phi = r \cos \theta, \; S_1 = r \sin \theta \cos \psi, \; S_2 = r \sin \theta \sin \psi.
\end{equation}
Substituting this parameterization to Eq. (\ref{Fintratrans}) and performing trigonometric transformations, we obtain
\begin{equation}
F = \alpha r^2 + \beta \frac{r^4}{4} \left[ 8 \left( \frac{1 - \frac{\cos^2 2 \psi}{8}}{1 - \frac{\cos ^2 2 \psi}{4}} \right) - 4 \left(1 - \frac{\cos ^2 2 \psi}{4} \right) \left( \cos 2\theta + \frac{\cos ^2 2 \psi}{4 - \cos^2 2\psi} \right)^2 \right].
\end{equation}
For $\beta<0$ the minimization requires maximizing the term in square brackets. The maximum is reached when
\begin{equation}
\cos 2\psi = \pm 1, \; \cos 2 \theta = - 1/3.
\end{equation}
Then, the expression  (\ref{Fintratrans})  for the minimum of energy reads
\begin{equation}
F = \alpha r^2 + 2 \beta r^4 \frac{1-\frac{1}{8}}{1-\frac{1}{4}} =  \alpha r^2 + \frac{7}{3} \beta r^4 .
\end{equation}
 The same free energy has been obtained in Ref. \cite{Shrock2010symmetry}
The resulting state breaks valley spin symmetry: one has either $\psi = 0$ (equivalent to $S_2 =0$), or $\psi = \frac{\pi}{2}$ (equivalent to $S_1=0$). Suppose we choose $S_2=0$. Then $S_1 = r \sqrt{\frac{2}{3}}$ and $\phi = \frac{r}{\sqrt{3}}$. Hence, the ground state is a coexistence state of $s^{\pm}$ valley (charge) order and spin order in one of the two valleys.

\section{Generic symmetry breaking patterns}

As SU(4) is a large symmetry group, its spontaneous breaking leads to a large ground state manifold. In the main text we considered perturbations around special states to compute the number of Goldstone modes. In this section, we obtain the generic patterns of SU(4) symmetry breaking in cases of first- and second-order phase transitions. The count of Goldstone modes follows naturally as the difference of generators of the full and residual symmetry groups
(Goldstone theorem).
 The analysis below is consistent with
  the one in
 Ref. \cite{Shrock2010symmetry}.

 \subsection{Free energy}

As we mentioned, the 15 order parameters relevant to us form a 15-dimensional adjoint representation of SU(4).
It is convenient in the following to
to consider it as a (reducible) representation of the O(4) subgroup of SU(4).  To this end, we represent the U(4) generators through the Dirac $\gamma$-matrices, or formally, the generators of the Clifford algebra $Cl_{4}(\mathbb{R})$, denoted as $\{\gamma_1,\gamma_2,\gamma_3,\gamma_4\}$. Further using well-known results in relativistic quantum field theory for transformation properties of $\gamma$-matrices, we group the 15 generators (labeled as $\{\Gamma^1,\cdots,\Gamma^{15}\}$) into four irreducible representations
of O(4):
\begin{align}
\textrm{vector:~}&\Gamma^{1,...,4}  = \gamma^{1,...,4}, \nonumber\\
\textrm{antisymmetric-tensor:~}&\Gamma^{5,...,10}= \gamma^{ij}\equiv \frac{[\gamma^i,\gamma^j]}{2i} \nonumber \\
\textrm{pseudo-vector:~}&\Gamma^{11,...14} =\gamma^{(1,...4)5}\equiv  i\gamma^{1,...4} \gamma^5, \nonumber\\
\textrm{pseudo-scalar:~}&\Gamma^{15} = \gamma^5 \equiv \gamma^1\gamma^2\gamma^3\gamma^4.
\label{eq:tensor}
\end{align}

When written in this basis, the matrix order parameter $\Phi$ can be expressed as
\be
\Phi = \phi_i\gamma^i + S_{ij} \gamma^{ij} + \psi_i \gamma^{i5}+ m_5\gamma^5,
\ee
it follows that $\vec \phi$ is a 4d vector, $\dvec S$ is a tensor, $\vec\psi$ is a pseudo-vector, and $m_5$ a pseudo-scalar. Incidentally, in this language the inner product of vectors is given by $\vec A\cdot \vec B=\Tr[AB]$ and the cross product is given by $\vec A \times \vec B = \frac{1}{2i}[A,B]$.

We can reexpress the matrix product $\Phi^2$ into irreducible representations of O(4), which will be useful
when calculating the trace for the free energy.
Using the properties that $(\gamma^i)^2 = \mathbb 1$ and $\gamma^5 \equiv \gamma^1\gamma^2\gamma^3\gamma^4$, we have
\be
\Phi^2= (\vec \phi^2+\vec\psi^2+ 2\dvec S:\dvec S + m_5^2) \mathbb{1}  + (\dvec S\times \vec \psi)_i\gamma^i  + (m_5 \dvec {\tilde {S}} + \vec\psi\times \vec \phi)_{ij} \gamma^{ij} + (\dvec S\times \vec \phi)_i\gamma^{i5}  + (\dvec S \times {\dvec{ S}}) \gamma^5,
\label{eq:phi2}
\ee
where $\dvec S :\dvec S\equiv S_{ij} S_{ij}$, and the cross products and dual tensors are defined as, e.g.,
\begin{align}
(\vec \psi \times \vec \phi)_{ij} = & \epsilon_{ijkl} \psi_{k} \phi_l, \nonumber\\
(\dvec S\times \vec \psi)_i = & \epsilon_{ijkl} S_{jk} \psi_l, \nonumber\\
\dvec S\times \dvec S = & \epsilon_{ijkl} S_{ij} S_{kl}, \nonumber\\
\dvec {\tilde {S}}_{ij} =& \epsilon_{ijkl} S_{kl}.
\end{align}

Defining the composite orders as
\begin{align}
\textrm{scalar:~}& R^2 = (\vec \phi^2+\vec\psi^2+ 2\dvec S:\dvec S + m_5^2),\nonumber\\
\textrm{vector:~}& \vec V = \dvec S\times \vec \psi,\nonumber\\
\textrm{antisymmetric tensor:~}& \dvec T =  m_5 \dvec {\tilde {S}} + \vec\psi\times \vec \phi \nonumber\\
\textrm{pseudo-vector:~}&\vec U =\dvec S\times \vec \phi \nonumber\\
\textrm{pseudo-scalar:~}& P = \dvec S \times {\dvec{ S}},
\end{align}
we have
\begin{align}
\Phi^2= R^2  \mathbb{1} + V_i\gamma^i + T_{ij}\gamma^{ij} + U_i \gamma^{i5} + P\gamma^5.
\label{eq:phi2}
\end{align}
According to the generators they are attached to, the composite field $R^2$
is a scalar, $\vec V$ a vector,  $\dvec T$ a tensor, $\vec U$ a pseudo-vector, and $P$ a pseudo-scalar.

The identification of O(4) irreducible representations
now helps greatly with organizing the terms in the free energy. At every order, only scalars can appear, which is enforced by the trace over the products of matrix order $\Phi$. From Eqs.~\eqref{eq:phi2}, using the
 defining property $\Tr[\Gamma^i]=0$ and
normalization convention
\be
\Tr[\Gamma^i \Gamma^j] =4\delta^{ij},
\ee
the SU(4) symmetric free energy can be expressed as
\begin{align}
F&= \frac{\alpha}{4}\Tr(\Phi^2) + \frac{3\gamma}{\sqrt2}\Tr(\Phi^3) + \frac{\beta}{4}\Tr(\Phi^4) + \cdots
\nonumber\\
&= \alpha R^2 + \beta R^4 +6\sqrt{2}\gamma(\vec V\cdot \vec\phi + \vec U \cdot \vec\psi +2 \dvec T:\dvec S + m_5 P) + \beta(\vec V^2+\vec U^2+ \dvec T^2 + P^2) + \cdots.
\label{eq:free}
\end{align}
Using input from our microscopic model, we set $\gamma=0$ hereafter. Note that purely from the perspective of SU(4) symmetry, an additional term $\sim [\Tr(\Phi^2)]^2$ is also allowed, but it is absent within our microscopic theory. Adding it would alter prefactor of the $R^4$ term.

\subsection{$\beta>0$, second-order transition}
In this case, the free energy is minimized by the vanishing of all non-scalar composite orders $\dvec T,\vec \psi, \vec\phi, P$.  This can indeed be achieved -- the simplest example is for $\Phi$ to be proportional to one of the generators $\Gamma^i$.
In this case, independent
 of the details of the ground state, from Eq.~\eqref{eq:phi2} we necessarily have
\be
\Phi^2 = R^2 \mathbb{1}.
\ee
This means that the operator $\Phi/|R|$ squares to the identity as any of the $\gamma$ matrices, and can be treated as  one of the generators of the Clifford algebra $Cl_{4}(\mathbb{R})$. In terms of the adjoint fields, such a ground state has an SO(4)$\otimes$ U(1) residual symmetry. To see this, let's choose a basis of the Clifford algebra such that
\be
\Phi/|R | = \tilde \gamma^1.
\ee
Among the 15 SU(4) generators, 7 commute with $\tilde \gamma^1$ and do not generate any variation in the ground state. They are $\tilde\gamma^1$ itself and $\tilde{\gamma}^{23}, \tilde{\gamma}^{24}, \tilde{\gamma}^{25}, \tilde{\gamma}^{34}, \tilde{\gamma}^{35}, \tilde{\gamma}^{45}$. Particularly, the latter six transform as a tensor under the SO(4) rotations among (2,3,4,5) axes just like $\gamma^{ij}$ for (1,2,3,4) axes in Eq.~\eqref{eq:tensor}. They  can be taken as  generators of SO(4) (different from the SO(4) subgroup of the O(4) we used to classify the 15 order parameters), which is a residual symmetry for the symmetry broken state with $\Phi\sim\tilde\gamma^1$. Moreover, $\tilde\gamma^1$ generates a U(1) group that is also a residual continuous symmetry. The symmetry breaking pattern in this situation can thus be written as
\be
\mathrm{SU(4)}\to {\mathrm{SO(4)}\otimes \mathrm{U(1)}}.
\label{eq:gold}
\ee
The number of Goldstone modes corresponds to the generators of the coset space $\mathrm{SU(4)}/ \left[{\mathrm{SO(4)}\otimes \mathrm{U(1)}}\right]$. By a simple count, there are $15-1-6=8$ Goldstone modes.

The residual symmetry can also be seen from the fermionic sector.  The term
\be
\Psi^\dagger \langle\Phi\rangle \Psi
\ee
obviously breaks SU(4), but not all its subgroups.  Since $\Phi^2/R^2 = \mathbb{1}$ and $\Tr[\Phi]=0$  the spectrum of $\Phi$ is
\be
\{R,R,-R,-R\}.
\ee
We clearly see that under such a mass term, the fermionic theory is still invariant under the symmetry group
\be
\mathrm{SU}(2)_+\otimes \mathrm{SU(2)}_- \otimes \mathrm{U}(1).
\label{eq:gold2}
\ee
 The two SU(2)'s act within the positive mass and negative mass sectors, and the U(1) is a relative phase rotation between the two sectors. Noting that $\mathrm{SO}(4)\cong \mathrm{SU(2)}\otimes \mathrm{SU(2)}/ \mathbb{Z}_2$, the residual symmetry in Eq.~\eqref{eq:gold2} is a double cover of that in Eq.~\eqref{eq:gold}, which is a result of the fermion parity symmetry acting trivially in the adjoint representation.

\subsection{$\beta<0$, first-order transition}

In the case of $\beta<0$, the phase transition is first-order and the ground state is obtained by maximizing the quartic term $F^{(4)}/|\beta|$. From Eq.~\eqref{eq:free}, for a given $R$, the ground state maximizes
\begin{align}
\vec V^2+\vec U^2+ \dvec T^2 + P^2 \equiv (\dvec S\times \vec \psi)^2+(\dvec S\times \vec \phi )^2 +(m_5 \dvec {\tilde {S}}+\vec\psi\times \vec \phi)^2+(\dvec S \times {\dvec{ S}})^2
\label{eq:align}
\end{align}
To analyze the properties of the ground state, a useful trick here is to use SU(4) symmetries to eliminate some of the terms in \eqref{eq:align}.  Specifically, one can perform an SU(4) transformation of the basis to eliminate the $m_5$ component using
\be
U = \exp \left(\frac{i\varphi \gamma^{15} }{2}\right),
\label{eq:su4}
\ee
where $\varphi$ is a rotation angle that satisfies  $\tan\varphi = m_5/\phi_1$. In this basis, one maximizes
\be
 \left(\vec\psi\times \vec \phi\right)^2+\left(\dvec S\times \vec \psi\right)^2+\left(\dvec S\times \vec \phi \right)^2+\left(\dvec S \times {\dvec{ S}}\right)^2
 \label{eq:align2}
\ee
under the constraint $\vec\phi^2+\vec\psi^2+2\dvec S:\dvec S=R^2$.

As can be readily verified, the maximum of  \eqref{eq:align2} corresponds to a configuration where $|\vec \psi|$, $|\vec \phi|$, and $|\dvec S|$ are all nonzero. By analogy with vector algebra, a
generic  saddle point of (\ref{eq:align2})
 is achieved when variations in each term of \eqref{eq:align2} vanish with respect to a change in the directions of $\vec \psi$, $\vec \phi$, and $\dvec{S}$.
 One
  way to satisfy this is to set  $\vec \psi \perp \vec \phi$.
  Without loss of generality, we take $\phi_3=\psi_4 \neq 0$, with all other components being zero. Simultaneously maximizing the second and
  the
   third term in (\ref{eq:align2}), we find that the
    nonzero components in $\dvec S$ are $S_{12}=-S_{21}$.
  Such a configuration is also a saddle point of the last term.
  We checked all other saddle points, such as configurations with $\vec\phi \| \vec\psi$,
  and found
  that the global maximum
  corresponds to the
  configuration that we just described, i.e., the one  with
\be
 \Phi = \phi( \gamma^3 + \gamma^{45} + \gamma^{12}).
 \ee
 Importantly,  we have shown that
 any
  ground state can be rotated to this configuration.

The generators corresponding to
 three non-vanishing components of the order parameters  commute.
In fact the three form a Cartan subalgebra of $\mathfrak{su}(4)$.
To determine
the fermionic spectrum with the mass term
$\Psi^\dagger \langle \Phi \rangle \Psi$, we go to their diagonal basis,  e.g.,
$\gamma^3=\textrm{diag}(1,1,-1,-1)$ and $\gamma^{12}
=\textrm{diag}(1,-1,1,-1)$. Then $\gamma_{45} = \textrm{diag}(-1,1,1,-1)$. It follows that the spectrum of $\Phi$ is given by
\be
\{R/\sqrt{3}, R/\sqrt{3}, R/\sqrt{3}, -\sqrt{3}R\}.
\ee
The threefold degeneracy in the ground state indicates that the symmetry breaking pattern is
\be
\mathrm{SU}(4) \to \mathrm{SU(3)}\otimes \mathrm{U}(1),
\ee the U(1) being a rotation in the relative phase between the positive mass and negative mass sectors. The number of Goldstone modes can be obtained by counting the generators of ${\mathrm{SU(4)}}/\left[{\mathrm{SU(3)}\otimes \mathrm{U(1)}}\right]$,
which is $15-8-1=6$.

%\bibliography{biblio}
\end{widetext}

\end{document}